\documentclass[aps,prx,reprint,superscriptaddress, floatfix]{revtex4-2}

\usepackage{graphicx,amsmath,amssymb,hyperref,physics} 

\usepackage[x11names]{xcolor}

\newcommand{\unit}[1]{\mathord{\mathrm{#1}}}

\begin{document}

\title{Sub-Doppler cooling, state preparation, and optical trapping of a triel atom}
\author{Putian Li}
\thanks{These authors contributed equally to this work.}
\affiliation{Centre for Quantum Technologies, National University of Singapore, 3 Science Drive 2, Singapore 117543}
\author{Xianquan Yu}
\thanks{These authors contributed equally to this work.}
\affiliation{Centre for Quantum Technologies, National University of Singapore, 3 Science Drive 2, Singapore 117543}
\affiliation{Duke Quantum Center, Duke University, Durham, NC 27701, USA}
\author{Seth Hew Peng Chew}
\affiliation{Department of Physics, National University of Singapore, 2 Science Drive 3, Singapore 117551}
\author{Jinchao Mo}
\affiliation{Centre for Quantum Technologies, National University of Singapore, 3 Science Drive 2, Singapore 117543}
\author{Tiangao Lu}
\affiliation{Centre for Quantum Technologies, National University of Singapore, 3 Science Drive 2, Singapore 117543}
\author{Travis L. Nicholson}
\email{travis.nicholson@duke.edu}
\affiliation{Centre for Quantum Technologies, National University of Singapore, 3 Science Drive 2, Singapore 117543}
\affiliation{Department of Physics, National University of Singapore, 2 Science Drive 3, Singapore 117551}
\affiliation{Duke Quantum Center, Duke University, Durham, NC 27701, USA}
\affiliation{Department of Physics, Duke University, Durham, NC 27708, USA}
\affiliation{Department of Electrical and Computer Engineering, Duke University, Durham, NC 27708, USA}

\begin{abstract}
Ultracold gases of atoms from Main Group III (Group 13) of the Periodic Table, also known as ``triel elements,'' have great potential for a new generation of quantum matter experiments. The first magneto-optical trap of a triel element (indium) was recently realized, but more progress is needed before a triel is ready for modern ultracold quantum science experiments in optical traps. Reaching this regime typically requires atoms that are cooled to the $10\,\unit{\mu K}$ level or below, prepared in pure quantum states, and confined in a laser field. Here we report the achievement of all three of these milestones in atomic indium. First, we perform polarization gradient cooling of an indium gas to $15\ \unit{\mu K}$. Second, we spin polarize the gas into a single hyperfine sublevel of either the $5P_{1/2}$ indium ground state or the $5P_{3/2}$ metastable state. Third, we trap indium in a 1064~nm optical lattice, achieving a 3 s trap lifetime. With these results, indium is now a candidate for a next generation quantum research platform.
\end{abstract}

\date{\today}
\maketitle

\section{Introduction}

Each time a new type of atom is cooled to ultralow temperatures, major advances in quantum science have followed. The first atom type to reach the ultracold regime, alkali metals, enabled breakthroughs such as quantum degenerate gases~\cite{Anderson1995,DeMarco1999}, spinor condensates~\cite{Stamper-Kurn2013}, and Hubbard model quantum simulations~\cite{Gross2017,Bohrdt2021}. The introduction of the second type, ultracold alkaline earth atoms, opened the door to advances like optical lattice clocks~\cite{Aeppli2024,McGrew2018} and SU$(N)$ interactions~\cite{Hofrichter2016,Taie2012,Sonderhouse2020}. Ultracold dipolar atoms~\cite{Chomaz2023}, with their anisotropic long-range interactions, gave rise to phenomena like emergent structural order~\cite{Chomaz2023,Tanzi2019,Bottcher2019,Chomaz2019,Petter2019} and dipole-stabilized phases~\cite{Kadau2016,Chomaz2016}.

The types of ultracold atoms realized so far lie in polar opposite regimes of angular momentum and magnetic interaction strength, leaving the middle ground largely unexplored. At one extreme, alkali and alkaline earth metals have zero orbital angular momentum ($L=0$) and few unpaired electrons, which results in weak magnetic dipole moments. At the other extreme, dipolar atoms exhibit large dipole moments due to high $L$, many unpaired electrons, or both. Meanwhile the rich intermediate landscape has tremendous potential for novel quantum science.

Among the atoms in this intermediate regime, the triel elements (Main Group III of the Periodic Table) stand out for their potential to realize exotic quantum many-body systems. Since these atoms have small dipole moments and $L=1$, their many-body interactions would be dominated by markedly anisotropic short-range scattering, a property that is novel for ultracold physics. In a Bose-condensed phase, this highly directional contact interaction could give rise to exotic phenomena such as anisotropic spinor physics. Similarly, the $F=4$ ground state of indium is predicted to support non-Abelian topological excitations under quantum-degenerate conditions~\cite{Serrano2023}. Indium also features both magnetic Feshbach resonances and an ultranarrow electronic clock transition compatible with stable laser technology. This combination, absent in alkalis and alkaline earths, would allow for an exceptional degree of quantum control, enabling tunable many-body interactions, spatially resolved coherent state manipulation, and probing with atomic clock precision. 

These features could support experiments at 10~$\mu$K level temperatures using many-body lattice systems probed with clock lasers, a platform previously explored in alkaline-earth atoms~\cite{Martin2013,Kolkowitz2017}. An indium-based version of such a system could access a broader range of many-body phenomena given its tunable interactions. At similar temperatures, triel atoms could also be loaded into optical tweezers for quantum simulations with single-particle control, following approaches demonstrated with alkalis and lanthanides. In both contexts, successful implementation typically requires temperatures of order $10\ \unit{\mu K}$, quantum-state purity, and optical trapping. 

In this work, we achieve all three of these milestones with indium: cooling two orders of magnitude below the Doppler limit, quantum state preparation with 90\% purity, and stable trapping of indium atoms in an optical lattice. Our techniques are broadly applicable to the triel elements, which share similar energy level structure as indium. These accomplishments set the stage for quantum science experiments with ultracold indium in optical lattices or tweezers. 

\begin{figure*}[t]
    \includegraphics[width=\linewidth]{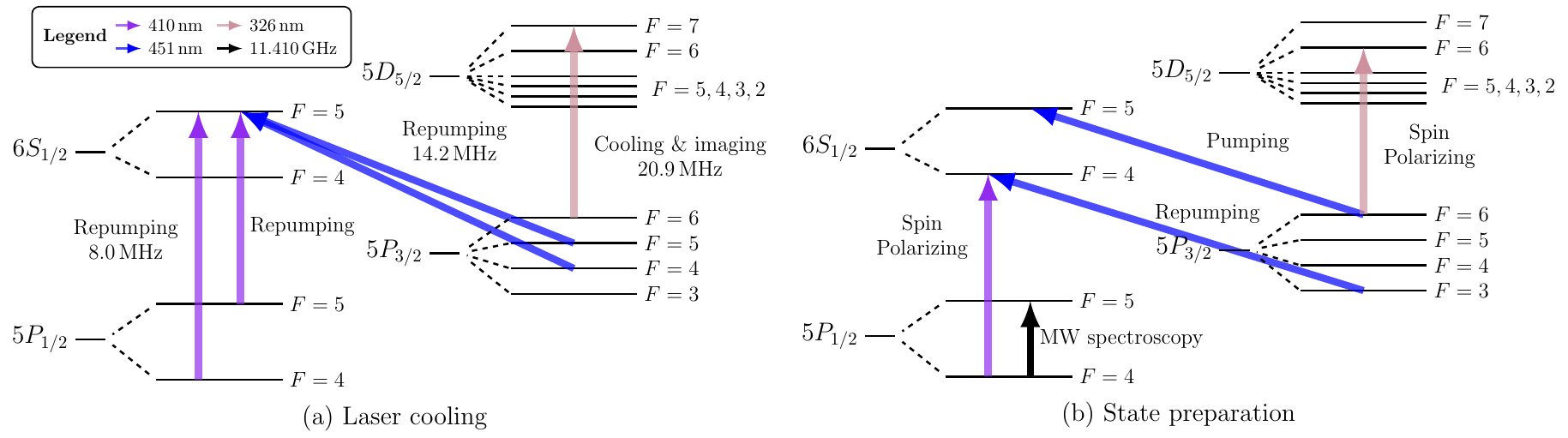}
    \caption{(a) Indium energy levels used for laser cooling. This scheme is presented in more detail in our previous work~\cite{Yu2022b}. The laser cooling transition is $\ket{5P_{3/2}, F=6} \rightarrow \ket{5D_{5/2}, F=7}$ throughout this manuscript. The frequencies in the figure are the natural linewidths of the respective transitions. (b) Atomic transitions used for spin polarization. Two spin polarization configurations are possible, allowing us to polarize atoms into either $\ket{5P_{1/2}, F=4, m_F=4}$ or $\ket{5P_{3/2}, F=6, m_F=6}$. The purity of spin polarization is determined with microwave spectroscopy.} \label{fig:levels}
\end{figure*}

\section{Sub-Doppler cooling}

We work with the most abundant indium isotope, $^{115}\mathrm{In}$. Its energy level diagram and the transitions used in this manuscript are shown in Fig.~\ref{fig:levels}. The first stage of our experiment involves collecting indium in a magneto-optical trap (MOT), which proceeds as follows. An indium atomic beam is generated by an effusion cell operating at $800\,^\circ\mathrm{C}$. The output is collimated by a microchannel array heated to $900\,^\circ\mathrm{C}$ to prevent clogging. The beam is then pumped into the $\ket{5P_{3/2},F=6}$ metastable cooling state using lasers at $410$ and $451\,\unit{nm}$~\cite{Yu2022}. The atoms are then decelerated to $70\,\unit{m/s}$ with a permanent magnet Zeeman slower. After that, they enter a MOT with a standard six-beam $\sigma^+ - \sigma^-$ configuration. The $\ket{5P_{3/2}, F=6} \rightarrow \ket{5D_{5/2}, F=7}$ transition at $326\,\unit{nm}$ is used for cooling in the Zeeman slower and MOT. In both of these stages, lasers driving the $\ket{5P_{3/2},F=4,5} \rightarrow \ket{6S_{1/2},F=5}$ transitions at $451\,\unit{nm}$ and the $\ket{5P_{1/2},F=4,5} \rightarrow \ket{6S_{1/2},F=5}$ transitions at $410\,\unit{nm}$ are present for repumping. With this setup, we have observed temperatures of order $1\,\unit{mK}$ (the Doppler limit is $500\,\unit{\mu K}$) while the atom number was as high as $1\times 10^9$~\cite{Yu2022b}.

\begin{figure*}[t]
    \includegraphics[width=\linewidth]{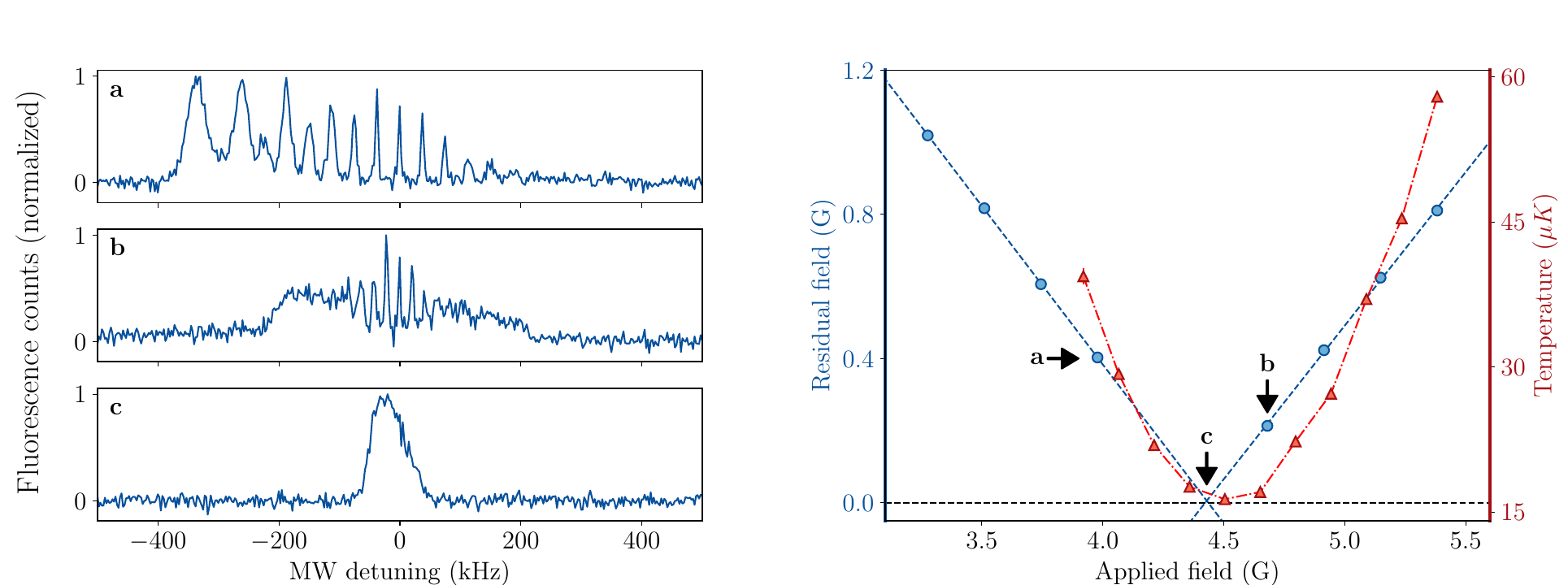}
    \caption{Left: Typical microwave spectra with decreasing external fields. Panel (c) shows the smallest Zeeman splitting we observed. A theoretical model of the spectrum explains the broadening of lines with larger $\abs{m_F}$ (the features that are farthest detuned from zero) as a result of a dynamic residual field~\cite{supplemental-material}. Right: The residual field (blue circles) and the measured temperature of the atoms (red triangles). The residual field was determined by the frequency difference between microwave resonances. The difference in applied field minima between the residual field and the temperature is because the two curves were measured at different dynamic residual field values~\cite{supplemental-material}.}\label{fig-pgc-field-zero}
\end{figure*}

\begin{figure*}
    \includegraphics[width=\textwidth]{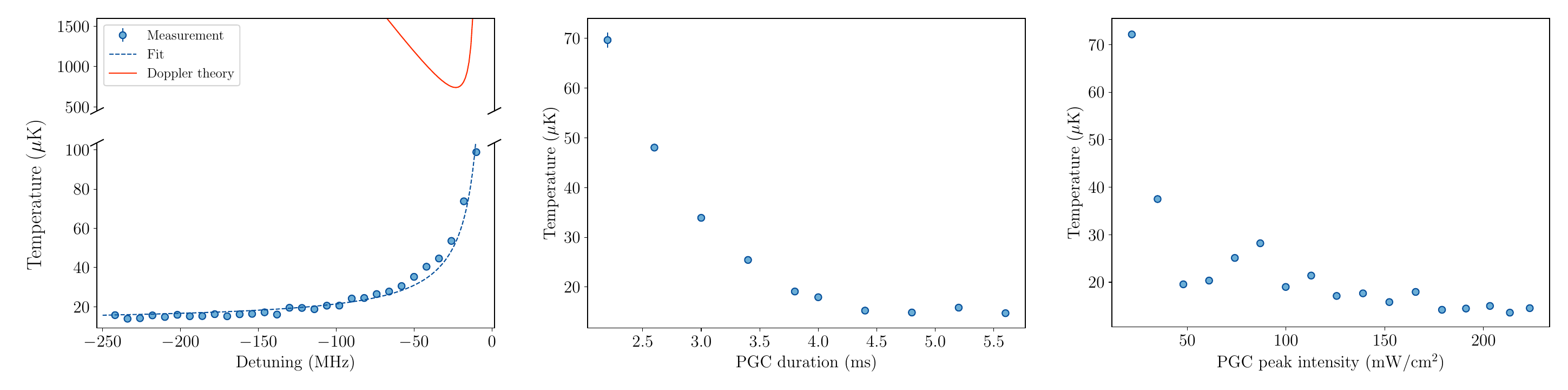}
    \caption{Characterization of polarization gradient cooling. Left:  Temperature vs. detuning, indicating sub-Doppler scaling in the temperature. The data is fit to the sub-Doppler cooling law $T = T_0 + T_1 (\Gamma/|\Delta|)$, where $T_0 = 7.0 \ \mathrm{\mu K}$ and $T_1 = 65.8 \ \mathrm{\mu K}$. In practice, the lowest temperature achieved is $15\,\mathrm{\mu K}$. Middle: Temperature vs. PGC pulse duration, which optimizes at $5\,\unit{ms}$. Right: Temperature vs. laser intensity, which shows the best performance at our maximum power.}\label{fig-pgc-char}
\end{figure*}

To achieve the goal of a $10\,\unit{\mu K}$ level temperature, we use polarization gradient cooling (PGC)~\cite{Dalibard1989,Salomon1991}, which works well when a cooling transition is in a hyperfine manifold with splittings much larger than the transition linewidth~\cite{Fernandes2012,Burchianti2014}, as is the case with indium. The primary concern with PGC is to ensure that any residual magnetic fields are well canceled, for nonzero bias fields lift the hyperfine sublevel degeneracy necessary for the PGC mechanism~\cite{Walhout1996}. Therefore, we develop a procedure to measure and compensate the residual magnetic field. We quantify this field by measuring the Zeeman splitting of the ground state hyperfine transition, $\ket{5P_{1/2}, F=4} \rightarrow \ket{5P_{1/2}, F=5}$ (at 11.410 GHz). The hyperfine $g$-factors for this transition predict that, in a bias magnetic field, it separates into 19 lines split by $93.3\,\unit{kHz/G}$ (calculated from the standard expressions for the $g$-factors, with the nuclear $g$-factor terms neglected). 

The sequence for microwave spectroscopy of the 11.410 GHz transition is as follows. After the initial MOT stage, atoms populate the lower-energy cooling state $\ket{5P_{3/2}, F=6}$. Atoms in this state are pumped into the hyperfine ground state $\ket{5P_{1/2}, F=4}$ with four lasers \cite{supplemental-material}. Three of these lasers have 451 nm wavelengths and drive the $\ket{5P_{3/2}, F=4,5,6} \rightarrow \ket{6S_{1/2}, F=5}$ transitions. The fourth laser has a $410\,\unit{nm}$ wavelength and drives the $\ket{5P_{1/2}, F=5} \rightarrow \ket{6S_{1/2}, F=5}$ transition. Since the MOT randomizes $m_F$ states~\cite{Tsyganok2018}, we expect the population to be distributed across many ground state hyperfine sublevels. We then drive the $\ket{5P_{1/2}, F=4} \rightarrow \ket{5P_{1/2}, F=5}$ hyperfine transition with a $800\,\unit{\mu s}$ microwave pulse. After the pulse is extinguished, we measure the population that completed the transition. To do this, we first apply a single $410\,\unit{nm}$ laser to drive atoms into $\ket{6S_{1/2}, F=5}$, which decays into the lower-energy cooling state with a branching ratio of 37\%. Then, we drive the cooling transition and collect its fluorescence, which is proportional to the population that made the microwave transition.

Using microwave spectroscopy, we find that the residual field has a static component, originating from magnetized objects in the lab or Earth's field, and a dynamic component due to eddy currents from shutting off the MOT coil. Both the static and dynamic components are large enough that we must compensate for them to obtain efficient PGC. To do this, we first add a delay before PGC to let the eddy currents partially dampen, and then we apply an empirically determined bias field ramp to compensate for the remaining residual field.

Fig.~\ref{fig-pgc-field-zero} illustrates typical microwave spectra for minimizing the residual field. The splitting between the microwave resonances varies as a function of the external magnetic field strength. The residual field is cancelled by measuring these spectra and adjusting the three bias fields until the splitting minimizes and the transitions are unresolved. With this technique, we determine the initial value of the compensating bias field ramp. This value reduces the residual field to below $50\,\mathrm{mG}$.

With this initial residual field cancellation, we implement a PGC sequence~\cite{supplemental-material}. First, the MOT is compressed for $5\,\unit{ms}$ by ramping up the quadrupole field, at which point the coil current and MOT laser power are shut off. Then we add the delay for dampening eddy currents before we pulse on the MOT lasers for PGC. By measuring the resulting temperatures, we empirically adjust the delay and compensating bias field ramp parameters during the PGC pulse for optimum cooling~\cite{supplemental-material}. 

We vary the duration, power, and detuning of the PGC pulse for best cooling (Fig.~\ref{fig-pgc-char}). The PGC pulse duration optimizes at $5\,\unit{ms}$, and the power optimizes at the maximum available intensity of $I/I_{sat} = 1.4$. The observed temperature $T$ is fit to the expected sub-Doppler cooling law of $T = T_0 + T_1(\Gamma/|\Delta|)$, where $\Delta$ is detuning, $\Gamma = 2\pi\times 20.9\,\unit{MHz}$ is the transition linewidth, and $T_0$ and $T_1$ are fit parameters. Although the fitted value of the minimum temperature is $T_0 = 7.0\ \unit{\mu K}$ (and $T_1 = 65.8\ \unit{\mu K}$), we do not typically achieve this in practice; therefore, we quote $15\ \unit{\mu K}$ as our minimum reliable temperature. This temperature, which is 30 times lower than the Doppler limit, is achieved when the detuning exceeds about $-150\,\unit{MHz}$ (or about $-7\Gamma$). Roughly half the atoms are lost during the PGC stage.

\section{State preparation}

\begin{figure*}[t]
    \includegraphics[width=\linewidth]{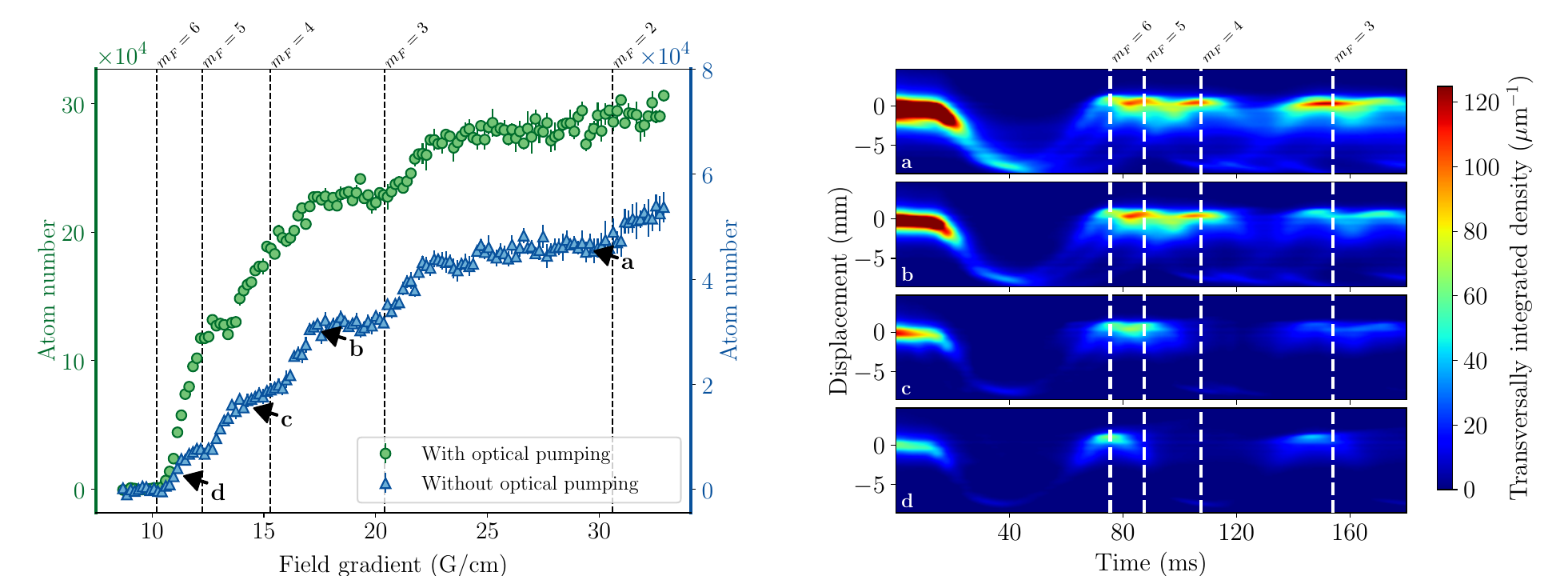}
    \caption{Left: Atom number as a function of the field gradient magnitude along the direction of gravity.  Green data (left y-ais) was taken with optical pumping; blue data (right y-axis) was taken without optical pumping. The step-like features~\cite{McCarron2011} occur when the force of gravity is equal to the confining force of the gradient for a given spin component. The atom number in the polarized component ($m_F=6$) is $1.3\times10^5$ ($6.8\times10^3$) with (without) optical pumping, so optical pumping boosts polarized atom number by a factor of 19. The annotations (a) through (d) indicate the points at which we measured the four plots [also labeled (a) through (d)] in the right panel. Right: In (a), several oscillation frequencies are apparent from the $m_F = 3,4,5,6$ spin components. In (d), only one oscillation frequency remains, indicating a spin polarized gas in the $m_F = 6$ state. The figure numbers (a) through (d) correspond to the measurement values (both field gradient and atom number) indicated in the left panel. The color bar represents the atomic density integrated along the $x$ and $y$ directions.}\label{fig-kick-spectrum}
\end{figure*}

Many important ultracold experiments require ensembles of atoms to be spin polarized into single hyperfine sublevels (as opposed to spin mixtures)~\cite{Aeppli2024,McGrew2018,Anderson1995,Stam2006}. We are interested in spin polarizing both the $5P_{1/2}$ ground state and the $5P_{3/2}$ metastable state, which has a predicted spontaneous lifetime of 10~s~\cite{Sahoo2011}. As previous work has shown, long-lived metastable electronic states have proved useful for studies of exotic interactions and quantum information processing~\cite{Hemmerich2006PRL,Daley2008,Gorshkov2009,Foss-Feig2010,Pucher2024,Unnikrishnan2024}. Here we realize spin polarization into the $\ket{5P_{3/2}, F = 6, m_F = 6}$ and $\ket{5P_{1/2}, F = 4, m_F = 4}$ states. 

\subsection{\texorpdfstring{$5P_{3/2}$}{5P\_3/2} state}
\label{subsec:5P3/2_prep}

State preparation schemes often involve using spin-dependent loss mechanisms to remove undesired spin components~\cite{Anderson1995,Stam2006} or optical pumping into stretched states~\cite{Aeppli2024,McGrew2018}. We perform state preparation in two phases, first using optical pumping and then with magnetic quadrupole trapping. At the center of a quadrupole field, the $z$ component of the force $\vec{\mathcal{F}}$ on an atom is
\begin{equation}
    \label{eqn:quadrupole_force}
    \mathcal{F}_z = -g_F m_F \mu_B \beta_z \mathrm{sign}(z) - m g,
\end{equation}
where $g_F = 1/3$ is the $g$-factor for the $\ket{5P_{3/2},F=6}$ state, $\mu_B$ is the Bohr magneton, the coordinate $z$ is antiparallel to gravity, $\beta_z$ is the magnitude of the magnetic field gradient at the center of the quadrupole trap (where $z=0$), and $m g$ is the gravitational force. The indium gas is loaded into the quadrupole trap after laser cooling, and this causes atoms with $m_F \leq 0$ to be lost (since they are not in trappable states). We then remove specific spin components by decreasing $\beta_z$. Note that when $\beta_z$ is reduced, Eqn. \ref{eqn:quadrupole_force} shows that all atoms with $m_F$ such that $-g_F \mu_B m_F\beta_z < mg$ will be overpowered by gravity and lost. Therefore, we lower $\beta_z$ until only atoms in the $m_F = 6$ stretched state remain (Fig.~\ref{fig-kick-spectrum}).

The purification scheme must be calibrated to determine the coil current at which all $m_F < 6$ spin states are ejected. We perform this measurement by ramping the magnitude of the gradient down to a low value $\beta_{min}$, waiting for untrappable spin components to drop away, and then ramping back up for strong confinement~\cite{supplemental-material}. Then we trigger a sudden shift of the quadrupole trap center by pulsing on a bias field in the $z$ direction. The shift causes the atoms to oscillate with a frequency related to the trap depth, which is proportional to $m_F$. With this frequency, we can identify different spin components from their oscillation frequencies (Fig.~\ref{fig-kick-spectrum}). By varying $\beta_{min}$, we can identify regimes where different spin components are present. The oscillations are well described by a known classical model~\cite{supplemental-material,Gomer1997eov}.

This quadrupole trap polarization scheme is lossy since it involves the ejection of many spin components. To mitigate this loss, we add a stage of optical pumping on the $\ket{5P_{3/2}, F=6} \rightarrow \ket{5D_{5/2}, F=6}$ transition with a $\sigma^+$-polarized laser (Fig.~\ref{fig:levels}b) before the gradient is ramped. During this pumping stage, we apply a $-16\,\mathrm{G}$ bias field to define the quantization axis. We find that the application of the spin polarization laser increases the polarized atom number by a factor of 19 (Fig. \ref{fig-kick-spectrum}, left).

\subsection{\texorpdfstring{$5P_{1/2}$}{5P\_1/2} state}

\begin{figure*}
    \includegraphics[width=\linewidth]{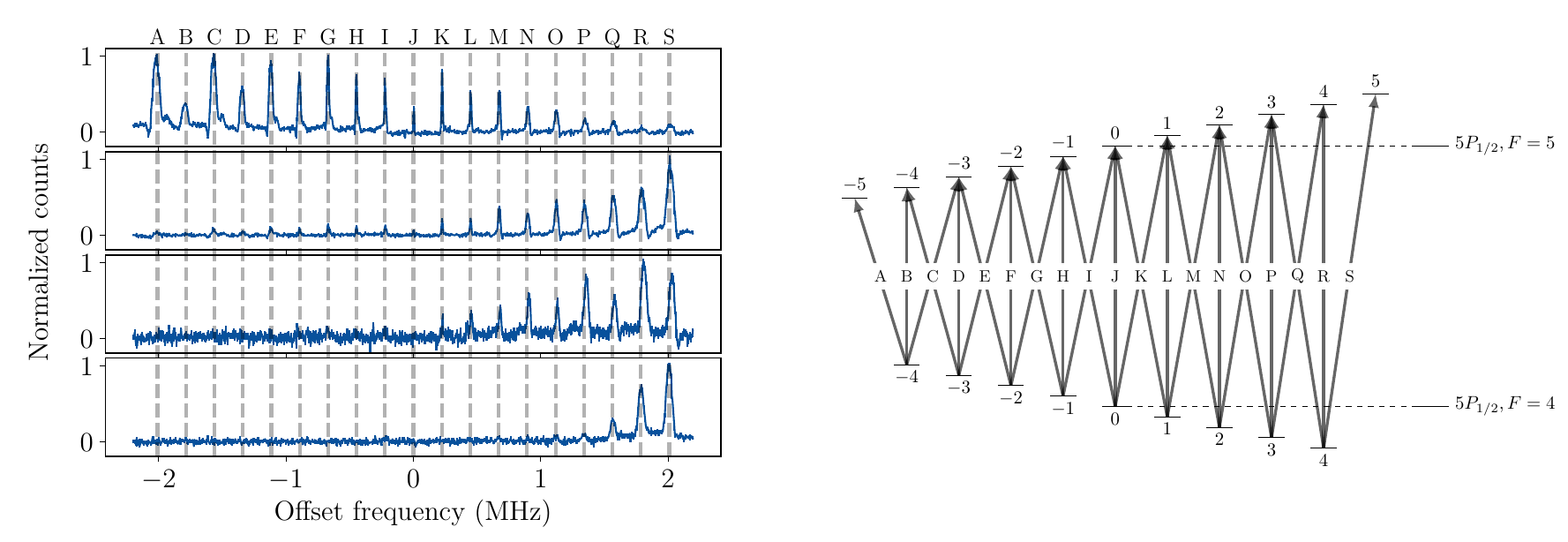}
    \caption{LEFT: Microwave spectra at different stages of the $5P_{1/2}$ spin polarization sequence. The top plot shows the microwave spectrum before spin polarization is applied, indicating many spin components. The second plot down shows the $5P_{1/2}$ population after spin polarization using the $\ket{5P_{3/2},F=6} \rightarrow \ket{5D_{5/2},F=6}$ transition. The third plot down shows the effect of holding atoms in the quadrupole trap to remove spin components with $m_F \leq 0$. The bottom plot shows the microwave spectrum after spin polarization with a $\sigma^+$-polarized 410 nm laser. The three prominent peaks are expected when the atoms are polarized in the target state, and they imply quantum state purity at the 90\% level. RIGHT: The transitions corresponding to those labeled in the observed microwave spectra.} \label{fig:fig-pump-spectrum}
\end{figure*}

The $g$-factor corresponding to $\ket{5P_{1/2},F=4}$ is five times smaller in magnitude than that of $\ket{5P_{3/2},F=6}$. Due to this weaker trap depth, our system is not currently capable of spin polarizing $\ket{5P_{1/2},F=4}$ using the quadrupole trap method employed for the $5P_{3/2}$ state. 

Therefore, we perform the following spin polarization sequence~\cite{supplemental-material}. First we polarize the laser cooled atoms using $\ket{5P_{3/2}, F=6} \rightarrow \ket{5D_{5/2}, F=6}$, and then we hold them in the quadrupole trap to remove residual $m_F\leq 0$ spin components. We then apply six lasers. The first of these lasers has a 451 nm wavelength and drives $\ket{5P_{3/2},F=6} \rightarrow \ket{6S_{1/2},F=5}$ to remove atoms from the metastable state. The second laser (at 410 nm and $\sigma^+$ polarization) drives $\ket{5P_{1/2},F=4} \rightarrow \ket{6S_{1/2},F=4}$, thereby polarizing atoms into $\ket{5P_{1/2},F=4,m_F=4}$. The third laser (at 451 nm) drives $\ket{5P_{3/2},F=3} \rightarrow \ket{6S_{1/2},F=4}$ to control leaks in the 410 nm spin polarizing transition. The remaining lasers are repumpers used in laser cooling, with two at 451 nm driving $\ket{5P_{3/2},F=4,5} \rightarrow \ket{6S_{1/2},F=5}$ and one at 410 nm driving $\ket{5P_{1/2},F=5} \rightarrow \ket{6S_{1/2},F=5}$.

The weaker quadrupole trap for $5P_{1/2}$ means we cannot determine the hyperfine sublevel populations with the quenching scheme used in $5P_{3/2}$. Therefore, we infer these populations with microwave spectroscopy of the $11.410\,\unit{GHz}$ ground state hyperfine transition $\ket{5P_{1/2},F=4} \rightarrow \ket{5P_{1/2},F=5}$. To identify which $m_F$ states are populated, we apply a 2.4~G magnetic field to fully split the microwave resonances (Fig.~\ref{fig:fig-pump-spectrum}). We evaluate different stages of the spin polarization sequence by pumping all atoms into the ground state and performing microwave spectroscopy to investigate the spin purity (Fig.~\ref{fig:fig-pump-spectrum}). When this sequence is complete, we infer from the spectrum that the atoms are polarized at the 90\% level~\cite{supplemental-material}.

\section{Trapping in an optical lattice}

\begin{figure*}[t]
    \includegraphics[width=\linewidth]{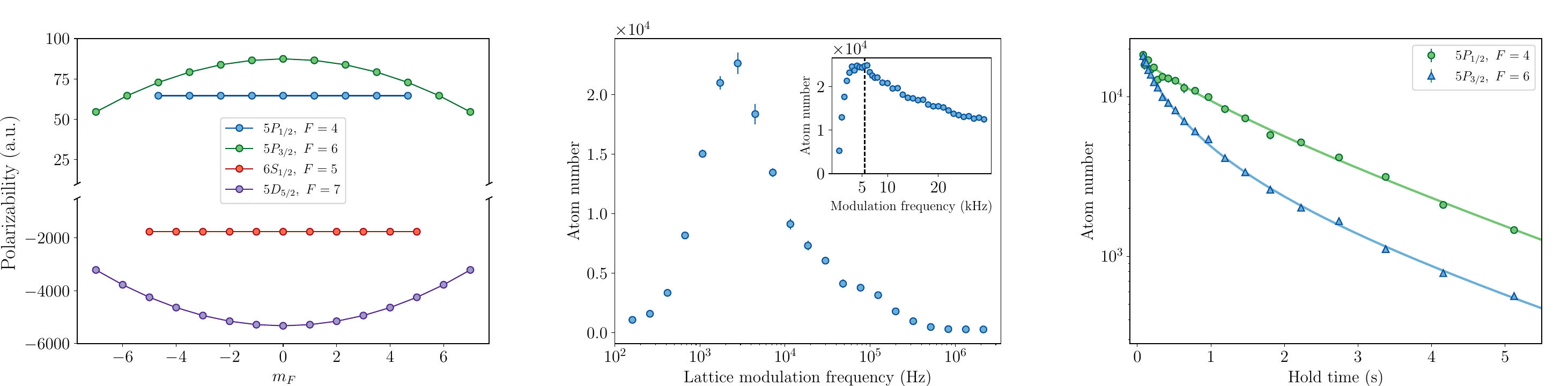}
    \caption{Optical trapping of indium in an optical lattice. Left: The dynamic polarizability of states relevant to loading indium into a lattice. The calculation is for a 1064~nm laser wavelength. The ground state $\ket{5P_{1/2}, F=4}$ and metastable state $\ket{5P_{3/2}, F=6}$ have positive polarizabilities while the repumping state $\ket{6S_{1/2}, F=5}$ and the cooling state $\ket{5D_{5/2}, F=7}$ have large and negative polarizabilities, causing inefficient trap loading.  Middle: Loaded atom number vs intensity modulation frequency. The inset shows the optimum modulation frequency of $5.5\,\mathrm{kHz}$. Right: Atom number vs lattice hold time for different atomic states. A substantial two-body loss feature is apparent for $\ket{5P_{3/2}, F=6}$. } \label{fig-ol}
\end{figure*}

With the achievement of low temperatures and spin polarization, we turn our attention to trapping in a one-dimensional (1D) optical lattice. This trapping potential has been used successfully in quantum simulations~\cite{Gross2017} and precision measurements~\cite{Aeppli2024,McGrew2018,Pezze2018}. We generate the lattice with a $1064\,\unit{nm}$ laser since there is a wide variety of high-powered, spectrally narrow sources available at this wavelength. To confirm good experimental performance at $1064\,\unit{nm}$, we calculate the indium polarizability using experimental atomic transition data~\cite{nist-database} when it is available and theoretical data~\cite{Safronova2007-all-transitions,Safronova2013-core} when it is not. 

Considering the case where the optical trap is linearly polarized along the quantization axis, we determine the hyperfine ground state polarizability at 1064 nm to be 65 a.u. (Fig.~\ref{fig-ol})~\cite{supplemental-material}. This value is small compared to trap wavelength polarizabilities of alkali metals, but it is still amenable to optical trapping with commercial laser sources. We also find that the $\ket{5P_{3/2},F=6}$ lower-energy cooling state polarizability is similar in magnitude to that of the ground state, and it has a substantial tensor polarizability term, which imparts a non-negligible $m_F$ dependence to the trap depth. For the $\ket{5D_{5/2},F=7}$ cooling state and the $\ket{6S_{1/2},F=5}$ repumping state, the polarizability is large in magnitude and negative, resulting in large Stark shifts and anti-trapping at 1064~nm. 

These effects cause inefficient trap loading, since the loading phase typically involves overlap between the laser cooling and optical trapping stages of an experimental sequence. To mitigate this inefficiency, trap intensity modulation has been successful~\cite{Hutzler2017,Phelps2020,Aliyu2021,Brown2019}. In this approach, one rapidly switches the trap intensity off and on, while the cooling light is switched $180^{\circ}$ out of phase with the trap~\cite{supplemental-material}. Previous work observed that loading is most efficient when switching frequencies are greater than the trap frequencies~\cite{Hutzler2017,Phelps2020,Aliyu2021,Brown2019}.

Using this modulation technique, we find that the optimal switching frequency for lattice loading is $5.5\,\mathrm{kHz}$ (Fig.~\ref{fig-ol}). A lattice potential model accounting for retroreflection mismatch predicts trap frequencies of $1.9\,\mathrm{kHz}$ ($247\,\mathrm{kHz}$) for the radial (axial) directions \cite{supplemental-material}. The switching frequency is well above the radial frequency but considerably less than the axial frequency. We note that while similar anti-trapping concerns arise in tweezer arrays, additional loading challenges beyond intensity modulation would also be present in such a system.

Loss mechanisms present in optically trapped gases include heating from off-resonant trap light scattering or inelastic collisions~\cite{Kuppens2000,Grimm2000}. To ensure that the atomic population is stable in the trap, we measure the in-trap atom number loss. Fitting this data to a model that contains both one- and two-body loss mechanisms~\cite{Yu2022b}, we observe a one-body trap lifetime of 2.7(2) s in the hyperfine ground state. For $\ket{5P_{3/2},F=6}$, the one-body lifetime is 3.0(1) s, and substantial two-body loss is apparent. Metastable state two-body loss in an optical trap is often caused by state-changing collisions~\cite{Grimm2000,Uetake2012}. More details on these decay rates and fits can be found within the Supplemental Material \cite{supplemental-material}.

\section{Conclusion}
With the realization of a $15\,\mu\mathrm{K}$ temperature, quantum state purity at the 90\% level, and stable optical lattice trapping, indium is now capable of optically trapped quantum science experiments. Both the $5P_{1/2}$ ground state and $5P_{3/2}$ metastable state can be used in future quantum science experiments. Potential experiments involve utilizing the $5P_{1/2} \rightarrow 5P_{3/2}$ indium clock transition, trapping in tweezers for materials simulations and quantum information~\cite{Jain2024}, evaporative cooling to a novel spinor BEC, and more. Additionally, synthetic spin-orbit coupling without heating is achievable when $L>0$~\cite{Burdick2016}, suggesting that triel elements could support spin-orbit-coupled spinor gases~\cite{Wang2010,Eustice2020}.

\hspace{1mm}

\section*{Acknowledgments}
This research is supported by the National Research Foundation, Singapore and A*STAR under its CQT Bridging Grant, as well as its Quantum Engineering Programme (Award No. NRF2022-QEP2-02-P15). It was also supported by a Duke University startup fund. We thank Paul Julienne for useful theoretical discussions and Huanqian Loh for reviewing the manuscript.

\nocite{*}
\bibliography{main}
\end{document}

\title{Sub-Doppler cooling, state preparation, and optical trapping of a triel atom: Supplemental Material}
\author{Putian Li}
\thanks{These authors contributed equally to this work.}
\affiliation{Centre for Quantum Technologies, National University of Singapore, 3 Science Drive 2, Singapore 117543}
\author{Xianquan Yu}
\thanks{These authors contributed equally to this work.}
\affiliation{Centre for Quantum Technologies, National University of Singapore, 3 Science Drive 2, Singapore 117543}
\affiliation{Duke Quantum Center, Duke University, Durham, NC 27701, USA}
\author{Seth Hew Peng Chew}
\affiliation{Department of Physics, National University of Singapore, 2 Science Drive 3, Singapore 117551}
\author{Jinchao Mo}
\affiliation{Centre for Quantum Technologies, National University of Singapore, 3 Science Drive 2, Singapore 117543}
\author{Tiangao Lu}
\affiliation{Centre for Quantum Technologies, National University of Singapore, 3 Science Drive 2, Singapore 117543}
\author{Travis L. Nicholson}
\email{travis.nicholson@duke.edu}
\affiliation{Centre for Quantum Technologies, National University of Singapore, 3 Science Drive 2, Singapore 117543}
\affiliation{Department of Physics, National University of Singapore, 2 Science Drive 3, Singapore 117551}
\affiliation{Duke Quantum Center, Duke University, Durham, NC 27701, USA}
\affiliation{Department of Physics, Duke University, Durham, NC 27708, USA}
\affiliation{Department of Electrical and Computer Engineering, Duke University, Durham, NC 27708, USA}
\date{\today}
\maketitle

\section{Polarization gradient cooling}\label{sec:app:pgc}

\subsection{Time sequencing}\label{subsec:pgc-time-sequence}
Polarization gradient cooling (PGC) requires that we time sequence the detuning, magnetic quadrupole field gradient, cooling laser intensity, and bias field. Here we describe the time sequencing. 

After MOT cooling was complete, we shut off the 326 nm cooling lasers and the quadrupole field. For the initial observation of PGC, we included a 9 ms delay after the quadrupole field shutoff was triggered. At this point, we performed microwave spectroscopy (Fig.~\ref*{fig-pgc-field-zero}, main text) and varied the bias coils to find the field minimum. The value of 9 ms was chosen to provide ample time for the dynamic component of the residual field to decay before the microwave pulse. Under these conditions, we measured the blue curve of main text Fig.~\ref*{fig-pgc-field-zero} (right). The bias field where the residual field minimizes was held constant throughout microwave spectroscopy, resulting in the optimum $z$ bias field value of 4.4 G shown in the main text.

Once a PGC signal was observed, further optimization revealed improvements by adjusting the delay and magnetic field compensation. We find that ramping the $z$ bias field provides improved final PGC temperatures. The red curve of main text Fig.~\ref*{fig-pgc-field-zero} (right) is measured under these optimized conditions. Here we describe the final optimized sequence. 

At the end of the MOT stage, we compress the MOT for 5 ms by ramping up the quadrupole field. The cooling lasers and quadrupole field are then switched off for 2 ms. We find that although making the delay longer than 2 ms better nulls the stray magnetic field, this results in substantial atom loss. During this 2 ms delay, we ramp the cooling laser detuning to $-200\,\unit{MHz}$ and the $z$-direction compensation coils to $5.2\,\unit{G}$. After the delay, we switch on the cooling light and trigger the dynamic magnetic field compensation scheme. 

The cooling light operates at an intensity of $I/I_\mathrm{sat} = 1.4$ for 6 ms to carry out PGC. This intensity corresponds to the full laser power, which we determine is best for decreasing the final temperature. Here $I_{sat}$ is calculated as described in our previous work~\cite{Yu2022b}. 

The dynamic magnetic field compensation scheme is as follows. For the first 3 ms, the $z$ compensation coil is linearly ramped from 5.2 G to 4.3 G. The ramp time and the final field value of 4.3 G are empirically chosen to minimize the final temperature after PGC. We observe no improvement by ramping the $x$ and $y$ bias fields, so these are held constant. We also attempt compensation waveforms that are more complex than linear ramps, such as two ramp periods with different slopes, but this does not affect the final temperature.

\begin{figure}    
    \includegraphics[width=.5\linewidth]{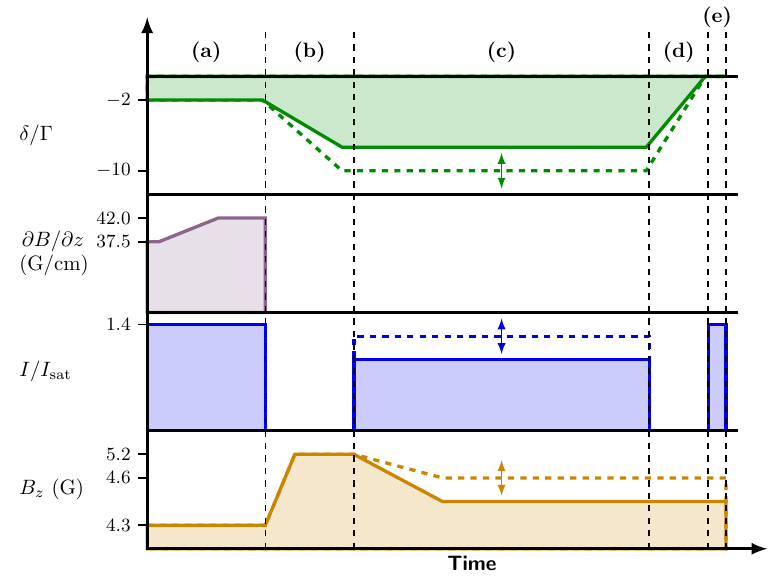}
    \caption{Time sequence for polarization gradient cooling. (a) MOT compression. (b) Delay for residual field to decay. (c) PGC cooling. (d) Free expansion for time-of-flight. (e) Fluorescence imaging.}\label{fig-pgc-sequence}
\end{figure}

\subsection{Microwave hardware}

Our microwave source is based on a PLL synthesizer operating at $11.410\,\mathrm{GHz}$. The synthesizer is amplified to 10 W with solid-state amplifier from RF Lambda. The resulting microwave field is directed to atomic cloud through a horn. 

\subsection{Schemes to pump and to image atoms before and after microwave spectroscopy}

\begin{figure}
    \centering
    \includegraphics[width=\linewidth]{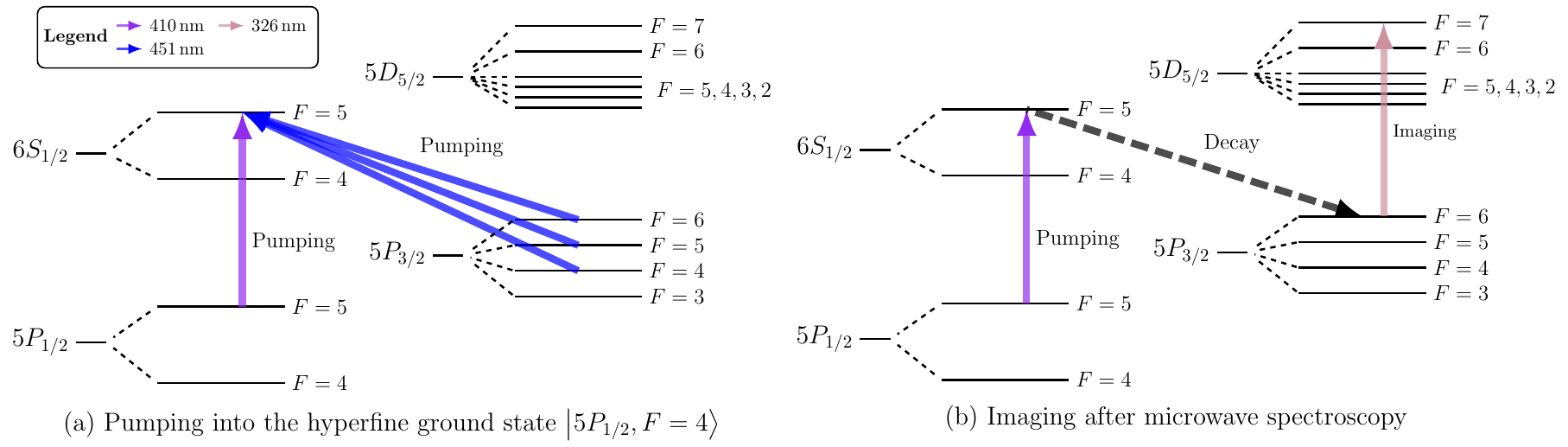}
    \caption{(a) The scheme to pump atoms from $\ket{5P_{3/2}, F=6}$ into $\ket{5P_{1/2},F=4}$. (b) The scheme to image atoms after microwave spectroscopy. }
    \label{fig:energy_level_mw}
\end{figure}

To pump atoms into the hyperfine ground state prior to microwave spectroscopy, we use one 410~nm laser and three at 451~nm (Fig. \ref{fig:energy_level_mw}(a)). Once microwave spectroscopy is performed, atoms need to be pumped back into the lower-energy cooling state for imaging. This second pumping scheme (Fig. \ref{fig:energy_level_mw}(b)) does not require many lasers due to the favorable branching ratio for atoms to decay into the lower-energy cooling state.

\subsection{Offset between magnetic field and temperature minima}
\label{subsec:field_offset}

In the right panel of Fig.~\ref*{fig-pgc-field-zero} of the main text, the residual field minimum and the minimum temperature occur at different applied fields. As mentioned above, the residual field minimum was measured with a 9 ms delay after the quadrupole field was switched off, and also the $z$ bias field was held constant. However, the temperature data was measured after optimization, which involved shortening this delay to 2 ms. This caused the PGC pulse to be measured in a different background magnetic field environment. 

Another caveat is that the $z$ bias field was ramped during the PGC pulse; however, only a single value of the compensation field is plotted on the x-axis of Fig.~\ref*{fig-pgc-field-zero} of the main text. Here we chose to plot the time averaged field across the PGC pulse.   

\section{State preparation} 

\subsection{Time sequence for \texorpdfstring{$5P_{3/2}$}{5P\_3/2} spin purification measurements}
Here we discuss the time sequence for the measurements presented in Fig.~\ref*{fig-kick-spectrum} of the main text. After loading the quadrupole trap from the MOT, which uses a magnetic field gradient of 42~G/cm, the purification procedure starts by linearly ramping the quadrupole field gradient down to the purification value in 50~ms. We choose a slow ramp duration to avoid causing center-of-mass motion in this stage. The gradient is maintained at the purification value for 250~ms to allow ample time for all unwanted spin components to be ejected from the trap. After this hold, we use fluorescence imaging to obtain the atom number. The resulting data is plotted in the left panel of Fig.~\ref*{fig-kick-spectrum} (main text). 

For the data in the right panel of Fig.~\ref*{fig-kick-spectrum} (main text), we again ramp the gradient from the MOT value of 42~G/cm to the purification value in 50~ms. Like before, the field is held at the purification value for 250~ms. After this hold, we increase the magnetic trap gradient to $\beta_z = 30\ \unit{G/cm}$ in 100 ms. At this gradient, all atoms with a magnetic quantum of $m_F\geq 3$ are trappable. Although using a higher quadrupole gradient at this stage would result in more spin components observed in Fig.~\ref*{fig-kick-spectrum} (main text, right panel), our calculations suggest that such a gradient would require a larger quench field than our experiment was designed for. After another 50~ms hold to allow transients to decay away, we pulse on the 10.3~G magnetic quench field in the $z$ direction. We then image the atoms as a function of the time after the quench field pulses on.

For the quench measurement, resolution between the $m_F$ states increases as a function of the quench field amplitude. However, if the quench field is too large, the cloud will be ejected out of the field of view of our camera. The 10.3~G field is chosen for the best resolution at which we could still observe the atomic cloud.

\subsection{Motion in the magnetic trap}
 
To investigate the atomic motion after the quench, consider a quadrupole magnetic field plus a bias field along the $z$ direction. We model the field as 
\begin{equation}
    \vec{B}=\left(-\frac{\beta_z x}{2},-\frac{\beta_z y}{2}, \beta_z z+B_q\right),
\end{equation}
where 
\begin{equation}
    \beta_z= \left. \frac{\partial B}{\partial z} \right|_{x=y=z=0}
\end{equation}
is the magnetic field gradient along the $z$ direction and $B_q$ is the quench field amplitude. The dynamics of this system are described in \cite{Gomer1997eov}. Ignoring motion in the $x-y$ plane, this problem reduces the model of \cite{Gomer1997eov} to the 1D semiclassical model referred to in the main text. The 1D force on a given $m_F$ state in the above-mentioned field and in the presence of gravity is 
\begin{equation}
\mathcal{F}_z(m_F)=g_Fm_F\mu_B\beta_z\,\mathrm{sign}\,(z-z_q)-mg\ ,\label{eq: MT_force}
\end{equation}
where $g_F = 1/3$ is the $g$-factor of $\ket{5P_{3/2}, F=6}$, $mg$ is the gravitational force, and $z_q = -B_q/\beta_z$ is the shift of the magnetic trap center from the pure quadrupole trap due to $B_q$. 

The trajectory of the oscillation is governed by Newton's equation of motion with the driving force of Eqn.~\ref{eq: MT_force}. The oscillation frequency $\nu(m_F)$ depends on the magnetic quantum number of the spin state as 
\begin{equation}
\nu(m_F) =\frac{1}{\sqrt{-8a_{+}(m_F)z_q}}\left[\frac{1}{a_{+}(m_F)}+\frac{1}{a_{-}(m_F)}\right]^{-1}.\label{eq:mt_osc}
\end{equation}
This expression only holds when $z_q<0$. Here $a_{\pm}(m_F) = g_F m_F\mu_B\beta_z/m\pm g$ is the acceleration of the atoms when they are away from the magnetic trap center, where the positive (negative) sign is for atoms above (below) the center. The trajectory of spin states is governed by the piecewise function
\begin{widetext}
    \begin{equation}
        z(t) = \begin{cases}
           -\Delta z_--\frac{1}{2}a_- [t-NT-(t_1+t_2)/2]^2, & t-t_1/2\in[NT,NT+t_2],\\
           -z_q - \frac{1}{2}a_+ (t-NT)^2, & \text{otherwise} 
       \end{cases}, \label{eq:MT-solution}  
       \end{equation}
\end{widetext}
where $\Delta z_-$ is the maximum deviation below the equilibrium position, $T=1/\nu(m_F)$ is the period of the oscillation, $N$ is the integer number of periods that have elapsed since $t=0$, and 
\begin{align}
    t_1 &= \sqrt{\frac{-8z_q}{a_+(m_F)}} \\
    t_2 &= \sqrt{\frac{-8z_qa_+(m_F)}{a_-^2(m_F)}}
\end{align}
are the times during each period that the atoms spend above (below) the equilibrium position. 

\begin{figure}
   \includegraphics[width=.5\linewidth]{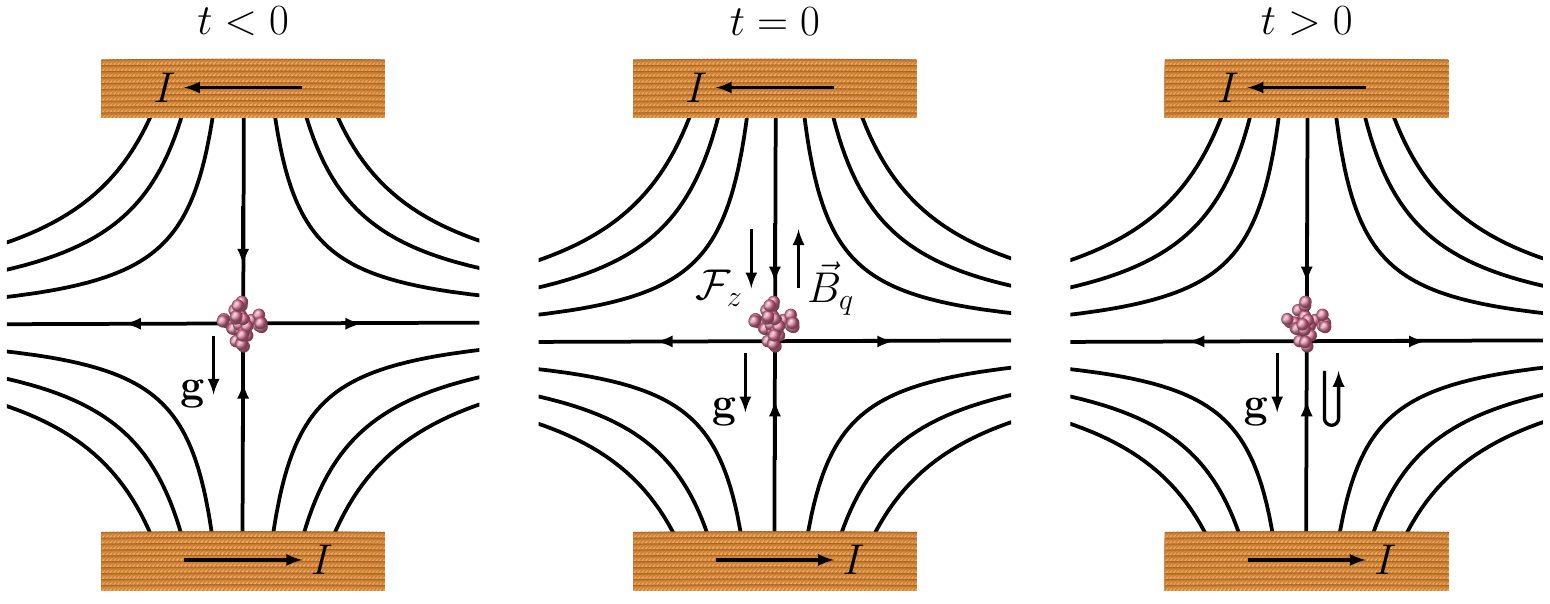}
     \caption{Time sequence to observe center-of-mass oscillations in the magnetic trap. LEFT: At $t<0$, atoms loaded into the magnetic quadrupole trap are in steady state in their equilibrium position. Middle: At $t=0$, pulsing on a bias field $B_q$ changes the center of the quadrupole trap, imparting a downward force $\mathcal{F}_{z}$. RIGHT: At $t>0$, oscillatory motion takes place.}
     \label{fig:kick_cartoon}
\end{figure}

\begin{figure}
    \includegraphics[width=.5\linewidth]{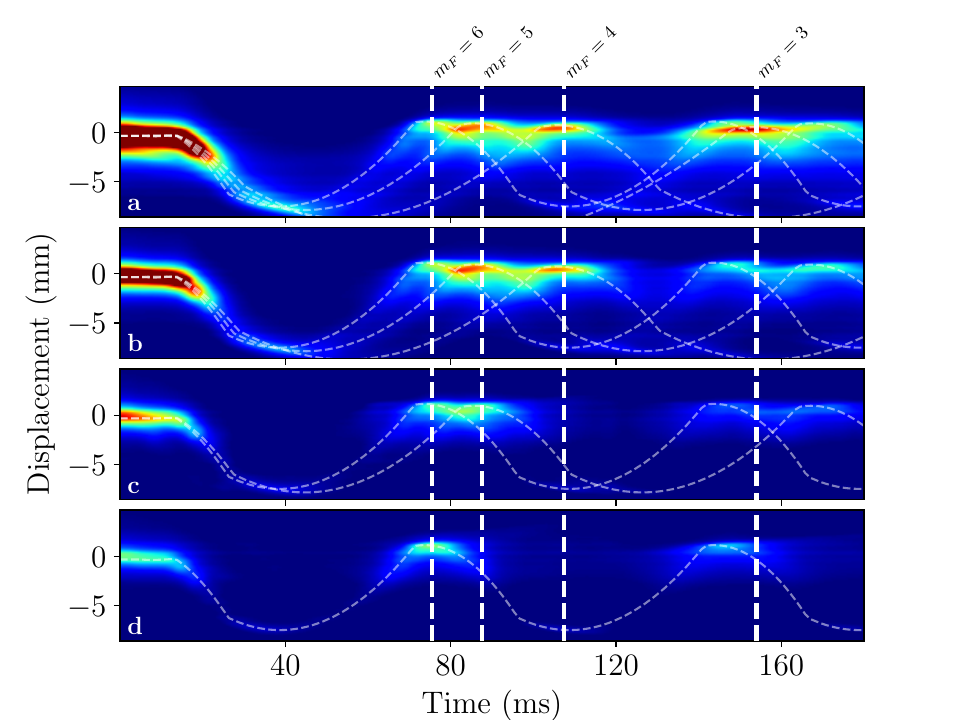}
    \caption{Theoretical quenching trajectories from Eqn.~\ref{eq:MT-solution} overlaid on Fig.~\ref*{fig-kick-spectrum} from the main text for $m_F=3,4,5,6$.}
    \label{fig:kick-with-fit} 
\end{figure}

In Fig.~\ref{fig:kick-with-fit}, the quench measurement data from Fig.~\ref*{fig-kick-spectrum} of the main text is compared with the theoretical trajectories from Eqn.~\ref{eq:MT-solution} for $m_F=3,4,5,6$. Despite that there appears to be good agreement, the theory is not fit to the data; rather, we have simply put independently measured experimental parameters into the model and overlaid the result with our data. The asymmetry of the theoretical oscillations occurs because the theory accounts for the 10 ms of free expansion that takes place before imaging. 

\subsection{Time Sequence for \texorpdfstring{$5P_{1/2}$}{5P\_1/2} optical pumping}

\begin{figure}
    \includegraphics[width=.5\linewidth]{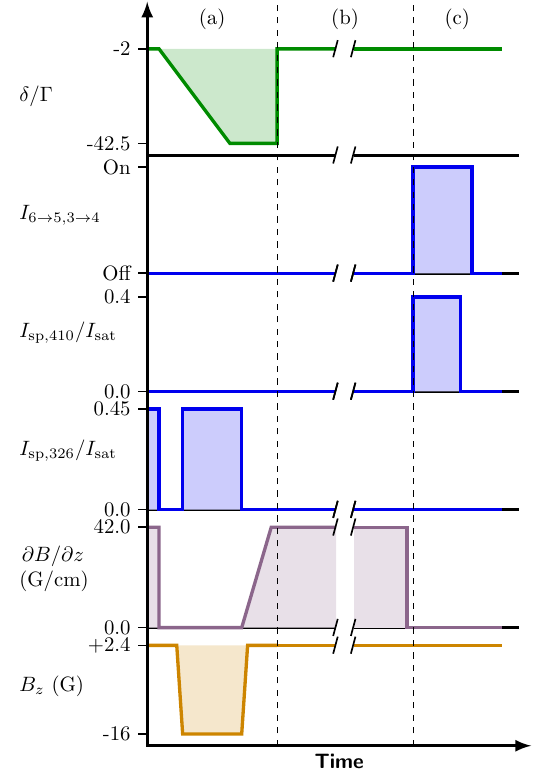}
    \caption{Time sequence for optical pumping. (a) The first stage of optical pumping (OP1). (b) Holding atoms in the quadrupole trap. (c) The second stage of optical pumping (OP2).}
    \label{fig:op}
\end{figure}

The state preparation process incorporates two optical pumping stages (OP1, OP2) and one magnetic trapping stage. The first stage of optical pumping (OP1) follows the MOT and lasts a total of 4.5~ms. It is also the initial stage to load atoms from the MOT into the magnetic quadrupole trap. Initially, we ramp the 326~nm polarizing laser frequency to $+207\ \unit{MHz}$ detuned from \OPtransition{} ($-41\Gamma$ detuned from the cooling transition) in 4.2~ms, which is limited by the frequency ramping speed of our laser system. Here, the maximum ramp speed is desirable because no atoms can be recaptured by the magnetic trap if the atoms are allowed to free-fall for more than 10~ms. Following this, we switch on the 326~nm polarizing laser (with a saturation parameter of $0.6$) for the final 1.5~ms to minimize heating and pushing effects due to near-resonant driving. This laser is circularly polarized, so to address the $\sigma^+$ transition, we align the polarizing laser to the quantization axis, which is determined by the $-16\ \unit{G}$ bias field along the $z$ direction. This field is turned on $500\,\unit{\mu s}$ before the light is shined and turned off at the end of the OP1 stage. The bias field must flip its sign because the $326\,\unit{nm}$ polarizing beam and the $410\,\unit{nm}$ polarizing beam (see below) are both $\sigma^+$ polarized but are counter-propagating. The value of $-16\,\mathrm{G}$ was optimized by maximizing the atoms loaded into the magnetic trap. The bias field before and after OP1 must not change because otherwise the magnetic field center would shift and atoms would be lost due to the motion discussed in the previous section.

Following OP1, atoms in the metastable state are pumped to $\ket{5P_{3/2}, F=6, m_F=6}$. Next, we hold the atomic sample in the magnetic quadrupole trap for 230~ms, ensuring all metastable state atoms with $m_F\leq 0$ are ejected. This hold time was chosen because we observed a decrease in the final purity when a shorter hold was applied. 

Finally, we employ OP2 for 2~ms. Here the quadrupole trap is shut off for 8 ms, and then the 410~nm spin polarization laser is turned on for the first 1.6~ms with a 2.4~G bias field. This 8 ms delay allows the eddy currents to decay, and we find that shortening this number decreases the final spin purity after OP2. Also, during OP2, we turn on all 451~nm MOT repumpers, the 410~nm repumper addressing $\ket{5P_{1/2},F=5}\rightarrow\ket{6S_{1/2},F=5}$, and additional repumpers addressing the $\ket{5P_{3/2},F=6}\rightarrow\ket{6S_{1/2},F=5}$ and $\ket{5P_{3/2},F=3}\rightarrow\ket{6S_{1/2},F=4}$ transitions. The 410~nm spin polarization laser is shut off earlier than the repumping beams to ensure that particles in the metastable state are fully pumped into the ground state. If ground state pumping is poor, this will result in residual metastable population that causes additional background counts in our microwave spectroscopy.

This background count issue is due to how we image the atoms. After a microwave pulse is applied, we expose the atoms to a 410~nm laser driving the $\ket{5P_{1/2},F=5}\rightarrow\ket{6S_{1/2},F=5}$ transition to populate the $\ket{5P_{3/2},F=6}$ state. Then, driving the cooling transition provides photon counts for imaging the atomic cloud. If there is population that is not part of the microwave sequence in the $5P_{3/2}$ manifold during imaging, it will provide background photon counts.

\subsection{Simulation of the optical pumping}

Many simple spin polarization schemes have been demonstrated, from straightforward optical pumping processes~\cite{Aeppli2024,McGrew2018} to single-stage spin filtering~\cite{Anderson1995,Stam2006}. Our approach is a multi-stage effort relying on both pumping and purification stages. To elucidate why this is so, we perform a rate equation simulation~\cite{Atoneche2017} of pumping to the $5P_{3/2}$ and $5P_{1/2}$ states.

For $5P_{3/2}$, the simulation proceeds as follows. Initially, population is equally distributed among the hyperfine sublevels of the $\ket{5P_{3/2}, F=6}$ state. The polarization of the optical pumping beam has an impurity of 0.1\%, and the repumping lasers are taken to be in crossed polarization (i.e., 50\% in the $\sigma^+$ component and 50\% in $\sigma^-$). Also, the saturation parameters are 0.2 for all beams, and the detunings are zero.

\begin{figure}
    \includegraphics[width=.5\linewidth]{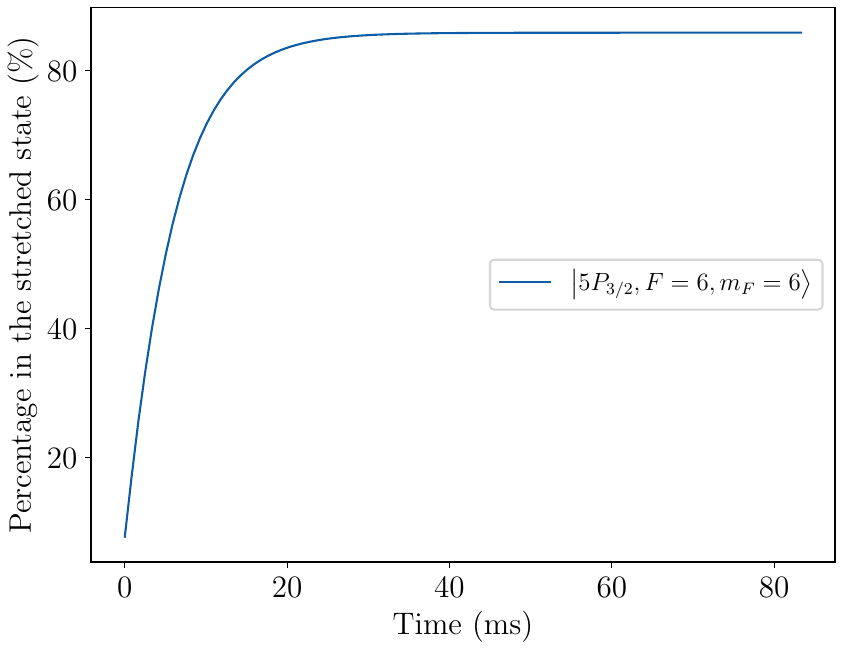}
    \caption{Simulation of optical pumping using the \OPtransition{} transition. The simulation parameters are described in the text. The final spin purity is 85.9\%.}\label{fig-sim-op}
\end{figure}

Fig.~\ref{fig-sim-op} shows the simulation result. We find that for these parameters, optical pumping can achieve a final spin purity of 85.9\% in tens of ms. After such a long time without trapping, there will be heavy atom number loss because these particles cannot be recaptured. Therefore, this optical pumping alone cannot generate a highly spin polarized ensemble. This is the reason we employ a spin purification stage with a quadrupole trap.

According to the simulation, a higher spin purity can be achieved considerably faster with much greater repumper power, which our experiment is not currently designed for. Also, in practice the 326~nm polarizing laser is detuned from resonance to prevent atom loss, which cannot be captured in this simulation (since it does not consider the momentum of the atoms).

For the $5P_{1/2}$ state, we set up the simulation in the following way. Atoms are initially polarized in the $\ket{5P_{3/2}, F=6, m_F=6}$ state. As before, the polarization of the optical pumping beam has a 0.1\% impurity, all pumping lasers have crossed polarization, and the saturation parameters are 0.2 for repumping beams. Additionally, the saturation parameter is 0.6 for the spin polarizing beam. All detunings are zero.

\begin{figure}
    \includegraphics[width=.5\linewidth]{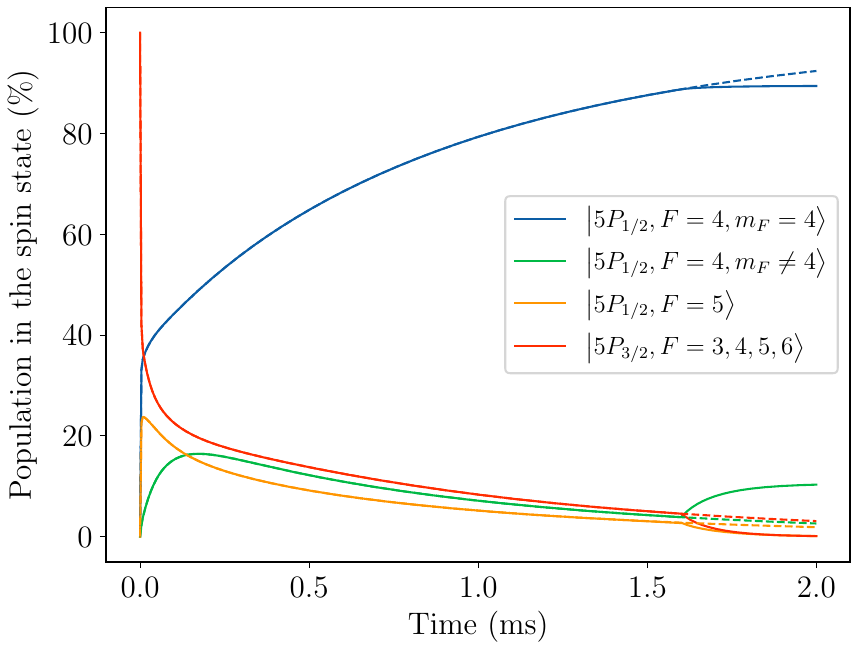}
    \caption{Simulation of the optical pumping using the \OPIItransition{} transition. The simulation parameters are described in the text. Shutting of the polarizing beam for the last $400\,\mu\mathrm{s}$ of the sequence keeps $5P_{3/2}$ state unpopulated at the end, reducing the background counts in microwave spectroscopy.}\label{fig-sim-op2}
\end{figure}

Fig.~\ref{fig-sim-op2} shows the simulation result for different spin states. Here the atoms polarize much faster because they are initially taken to be fully polarized into $\ket{5P_{3/2}, F=6, m_F=6}$. The solid lines represent the case where the OP2 polarizing beam is shut off for the last $400\,\unit{\mu s}$ of the sequence. The dashed lines show what would happen if the polarizing beam were kept on the entire sequence. We find that although turning the polarizing beam off at the end of the sequence reduces the final spin purity, it also reduces the number of atoms in the $5P_{3/2}$ state. As mentioned above, atoms in this state increase background counts in microwave spectroscopy, so shutting off the laser seems a worthy trade off.

\subsection{Microwave spectroscopy in a time varying magnetic field}\label{sec-microwave-atom-number}

In this section, we derive a model to explain the population dynamics in a time varying external magnetic field, solve it numerically, and apply it to the measured spectrum (Fig.~\ref*{fig:fig-pump-spectrum} of the main text) to extract the spin purity. As mentioned in Section~\ref{subsec:field_offset}, our system has a transient magnetic field. This originates from shutting off the quadrupole field used for spin purification.

The microwave spectroscopy can be modeled with the Lindblad master equation in the rotating-wave approximation, 
\begin{equation}
    \dot{\rho} = -i[H(t),\rho] + \mathcal{L}[\rho], \label{eq:spin-motion-full}
\end{equation} 
where the time-dependent Hamiltonian is
\begin{equation}
    H(t)=\frac12\begin{bmatrix}
        \Delta - \epsilon t & \Omega \\ \Omega & -\Delta + \epsilon t
    \end{bmatrix}.
\end{equation}
Here we work in the basis consisting of ground state $\ket{g}=\begin{bmatrix}
    1 \\ 0
\end{bmatrix}$ and excited state $\ket{e}=\begin{bmatrix}
    0 \\ 1
\end{bmatrix}$. Also, $\Delta$ is the tuning, $\Omega$ is the Rabi frequency, and 
\begin{equation}
    \epsilon = (g_e m_e - g_g m_g) \mu_B \pdv{B}{t}
\end{equation}
quantifies the transient field. $g_e$ ($g_g$) and $m_e$ ($m_g$) is the excited (ground) state $g$-factor and hyperfine magnetic quantum number, respectively. 

By performing microwave spectroscopy with a variable delay, we study how the microwave resonance frequency changes with time. We observe that the resonance frequency has a constant rate of change over the course of the microwave pulse. From this measurement, we infer that the rate of change of the field is 
\begin{equation}
    \frac{\partial B}{\partial t} = 70\,\unit{mG/ms}.
\end{equation}
To give an indication of how much this shifts the resonance, for $m_e = m_g = 4$, this corresponds to
\begin{equation}
    \epsilon = 52\,\unit{kHz/ms},
\end{equation}
which is a shift of about 42 kHz over the microwave pulse. 

The Lindbladian describing dephasing is
\begin{equation}
    \mathcal{L}[\rho] = \gamma (\sigma_z\rho \sigma_z-\rho). 
\end{equation}
Here $\sigma_z=\ket{g}\bra{g}-\ket{e}\bra{e}$ is the usual Pauli spin operator, and $\gamma$ is the dephasing rate. We have neglected spontaneous emission since the associated rate is very small for this type of transition. 

We consider two possible origins of $\gamma$: magnetic field noise (the effect of which would be proportional to $m_F$) and microwave field noise. We then extract the non-magnetic component by measuring the dephasing of Rabi oscillations on the $m_g = 0$ to $m_e = 0$ transition, which is first-order insensitive to magnetic shifts. Fitting the fringe visibility of the oscillations, we find the non-magnetic part of the dephasing rate to be
\begin{equation}
    \gamma_{\mathrm{non\,mag}} = 7.0(5)\,\unit{kHz}.
\end{equation}
Fitting the $m_g=4$ to $m_e=4$ lineshape, we find that the total dephasing rate is $\gamma \simeq 9\gamma_{\mathrm{non\,mag}}$. This indicates that, in this model, magnetic field noise dominates the dephasing.

\subsection{Extracting the spin purity from the observed microwave spectrum}
After the spin polarization stage, population has accumulated in the target stretched state $\ket{5P_{1/2},F=4,m_F=4}$, with some residual population in other states. Our aim is to determine what fraction of the population ended up in the stretched state.

To do this, we split the microwave states in a magnetic field and perform microwave spectroscopy, driving atoms into the excited hyperfine state $\ket{5P_{1/2},F=5}$. We then fluoresce the excited atoms and collect a portion of the resulting photon emission. Studying this as a function of the microwave frequency, we can identify $\pi$ transitions, as shown in Fig.~\ref*{fig:fig-pump-spectrum} of the main text. The measured $\pi$ transition lineshapes are fit to Voigt functions to extract the amplitudes, and these can be used to infer the spin purity. Although we have no a priori reason to expect the lineshapes will fit Voigt profiles, these are chosen because they can accommodate both homogeneous and inhomogeneous broadening, and they give better amplitude fits than other standard functions like the Lorentzian or the Gaussian. 

To find the spin purity with the amplitudes of the Voigt profile, we must note a few relationships between relevant quantities. The ground state atom number for a given hyperfine sublevel $N_g(m_F)$ is related to the corresponding number of excited state atoms $N_e(m_F)$ as
\begin{equation}
    N_e(m_F) = \kappa_{\mathrm{\mu W}}(m_F) N_g(m_F),
\end{equation}
where $\kappa_{\mathrm{\mu W}}(m_F)$ is the fraction of atoms driven by the microwave field from the ground state into the excited state. The excited state atom number is related to the collected fluorescence signal $\mathcal{S}(m_F)$ as
\begin{equation}
    \mathcal{S}(m_F) = \kappa_{sig}(m_F)N_e(m_F),
\end{equation}
where $\kappa_{sig}(m_F)$ quantifies the number of emitted photons collected by the detector. The fluorescence signal is related to the fitted Voigt profile amplitude $\mathcal{A}(m_F)$ as
\begin{equation}
    \mathcal{A}(m_F) = \kappa_{fit}(m_F)\mathcal{S}(m_F).
\end{equation}
Here $\kappa_{fit}(m_F)$ is one if the fit provides the exact microwave peak amplitude, and it can deviate from one if there are systematic errors associated with the fits. 

The hyperfine sublevel purity $\wp$ after spin polarization is
\begin{align}
    \wp &= \frac{N_g(4)}{\sum_{m_F=-4}^4 N_g(m_F)} \\
    &\simeq \frac{N_g(4)}{N_g(3)+N_g(4)} \\
    &= \frac{\frac{\mathcal{A}(4)}{\kappa_{\mathrm{\mu W}}(4)\kappa_{sig}(4)\kappa_{fit}(4)}}{\frac{\mathcal{A}(4)}{\kappa_{\mathrm{\mu W}}(4)\kappa_{sig}(4)\kappa_{fit}(4)} + \frac{\mathcal{A}(3)}{\kappa_{\mathrm{\mu W}}(3)\kappa_{sig}(3)\kappa_{fit}(3)}}.\label{eqn:spin_purity}
\end{align}
Here the approximation results from the fact that, after spin polarization, the only two $\pi$ transitions with statistically significant amplitudes are $\ket{m_F = 4} \rightarrow \ket{m_F = 4}$ and $\ket{m_F = 3} \rightarrow \ket{m_F = 3}$.

\begin{figure}
    \includegraphics[width=.5\linewidth]{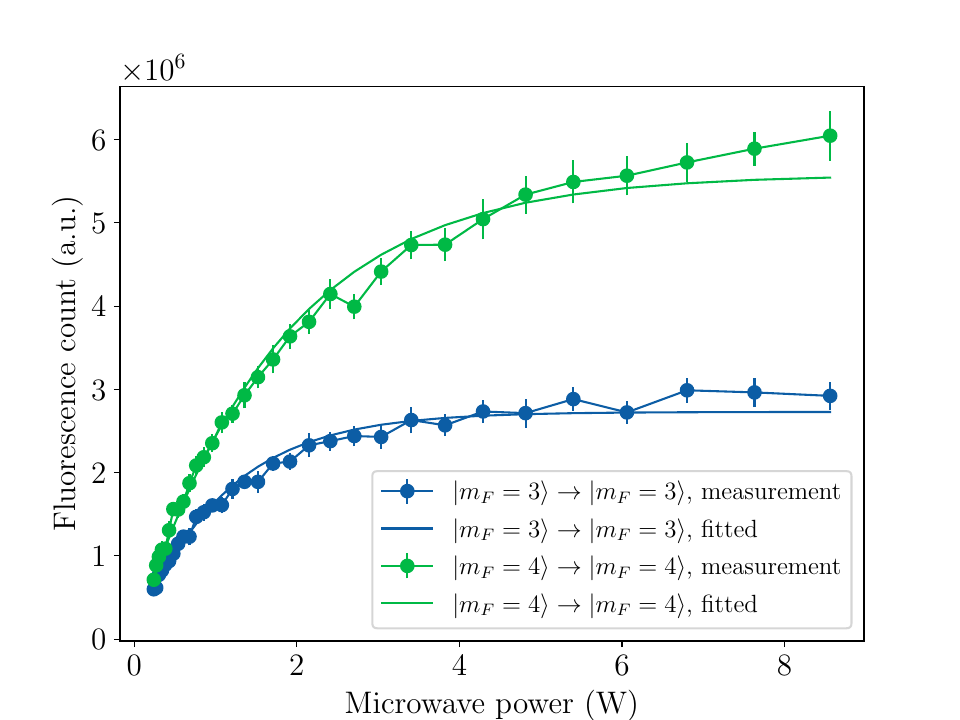}
    \caption{Peak fluorescence counts from $\pi$ transition resonances vs the microwave power. The $\ket{m_F=3}\rightarrow\ket{m_F=3}$ fit has a reduced chi square of $\chi_{red}^2=0.8$, and the $\ket{m_F=4}\rightarrow\ket{m_F=4}$ fit has $\chi_{red}^2=1.2$.}
    \label{fig:microwave_saturation}
\end{figure}

As shown in Eqn. \ref{eqn:spin_purity}, the proportionality constants $\kappa_{\mathrm{\mu W}}(m_F)$, $\kappa_{sig}(m_F)$, and $\kappa_{fit}(m_F)$ do not contribute to $\wp$ if they are independent of $m_F$. Although $\kappa_{\mathrm{\mu W}}(m_F)$ is generally dependent on $m_F$, it is equal to 1/2 when the transition is saturated. To ensure we are operating in the saturated regime, we vary the microwave power and measure the corresponding fluorescence on a transition resonance (Fig. \ref{fig:microwave_saturation}). This data is fit to our master equation model. The data does not show as much saturation as the model predicts; however, a 10\% level deviation from this condition results in a correction at only the 1\% level. Therefore, it is accurate within the 1\% level to treat the microwave spectroscopy as saturated, so we proceed with this.

To determine the $m_F$ dependence of $\kappa_{sig}(m_F)$, it is helpful to consider how imaging works in our system. Atoms in a particular $m_F$ state are pumped into $\ket{5P_{3/2},F=6}$ and then fluoresced by the cooling laser. Although atoms will depolarize out of their initial $m_F$ value during this pumping process, the initial $m_F$ state has some bearing over what the $m_F$ state ends up being when atoms are pumped into $\ket{5P_{3/2},F=6}$. In light of the well-known result in atomic physics that atoms decaying between two hyperfine levels will have decay rates that are independent of $m_F$ (since summing over different dipole emission pathways removes the $m_F$ dependence), $\kappa_{sig}(m_F)$ would also be $m_F$ independent.

\begin{figure}
    \includegraphics[width=\textwidth]{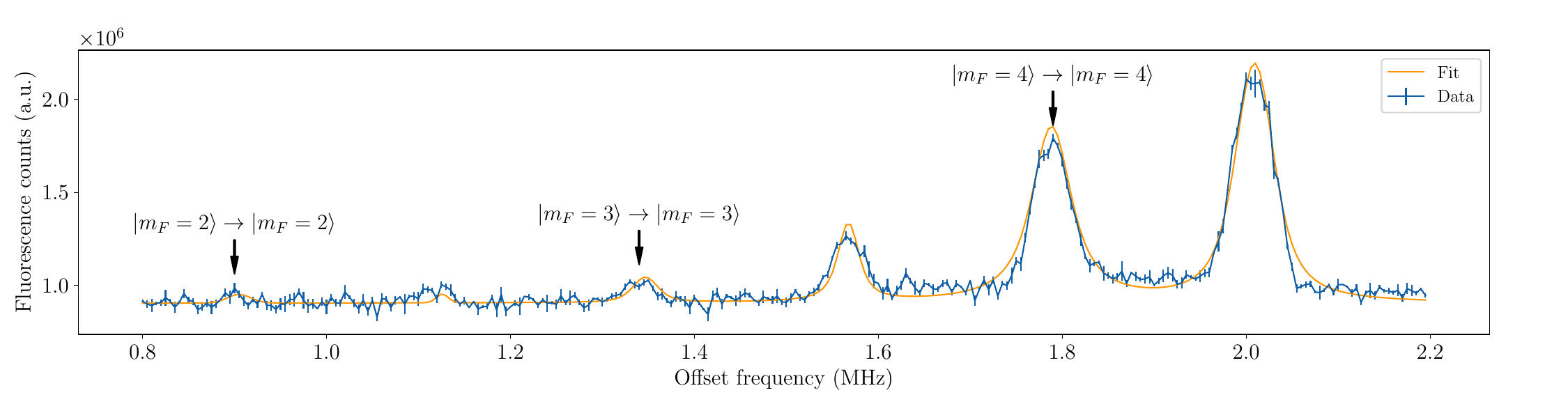}
    \caption{Fitting of the microwave spectrum. The microwave spectrum of the last $6$ transitions is fitted to a sum of $6$ Voigt profiles. The final spin purity is $\wp=87(4)\%$. }\label{fig:fitting_mw}
\end{figure}

Lastly, our numerical studies show that $\kappa_{fit}(m_F)$ is also independent of $m_F$. As Fig. \ref{fig:fitting_mw} shows, the Voigt profiles do not appear to give a great fit for the observed spectroscopic features. However, using our master equation theory (which better reproduces the qualitative lineshape features), we obtain an amplitude correction factor to properly scale the fitted peak height to the actual value. We find that this correction factor is robust against variations in Rabi frequency and other parameters, indicating that it is the same for each observed microwave feature. Therefore, we also treat $\kappa_{fit}(m_F)$ as independent of $m_F$.

Fitting the microwave spectroscopy data, we find that 
\begin{align}
    \mathcal{A}(4) &= 9.4(1)\times10^5 \\
    \mathcal{A}(3) &= 1.36(6)\times10^5,
\end{align}
resulting in the spin purity
\begin{align}
    \wp &\simeq \frac{\mathcal{A}(4)}{\mathcal{A}(3)+\mathcal{A}(4)} \\
    &= 87(4)\%.
\end{align}

\section{Optical lattice}
\subsection{Polarizability calculation}
The theoretical polarizabilities used in Fig. \ref*{fig-ol} of the main text were calculated using the formalism of ~\cite{LeKien2013}. This approach is based on the scalar-vector-tensor decomposition of the polarizability, 
\begin{equation}
\begin{aligned}
    \alpha&_{\ket{J,F,m_F}}(\omega,\vec{u})=\alpha_{\ket{J,F}}^s(\omega)-i\alpha^v_{\ket{J,F}}(\omega)\frac{[\vec{u}^*\cross\vec{u}]\cdot\vec{F}}{2F}\\ 
    &+\alpha_{\ket{J,F}}^t(\omega)\frac{3[(\vec{u}^*\cdot\vec{F})(\vec{u}\cdot\vec{F})+(\vec{u}\cdot\vec{F})(\vec{u}^*\cdot\vec{F})]-2\vec{F}^2}{2F(2F-1)}\ ,\label{eq: polarizability}
\end{aligned}
\end{equation}
where $\omega$ is the angular frequency of the laser field, and $\vec{u}$ is the Jones vector of the laser polarization. $\ket{J, F, m_F}$ labels the state of the atoms in the light field, and $\alpha_{\ket{J, F}}^s(\omega)$, $\alpha_{\ket{J, F}}^v(\omega)$, and $\alpha_{\ket{J, F}}^t(\omega)$ are the dynamical scalar, vector, tensor polarizabilities, respectively. These polarizabilities are defined as~\cite{LeKien2013}
\begin{align}
    \alpha^{s}_{nJF}&=\frac{1}{\sqrt{3(2J+1)}}\alpha_{nJ}^{(0)},\\
    \alpha^{v}_{nJF}&=(-1)^{J+I+F}\sqrt{\frac{2F(2F+1)}{F+1}}\begin{Bmatrix}F&1&F\\J&I&J\end{Bmatrix}\alpha^{(1)}_{nJ},\\
    \alpha^{t}_{nJF}&=-(-1)^{J+I+F}\sqrt{\frac{2F(2F-1)(2F+1)}{3(F+1)(2F+3)}}\begin{Bmatrix}F&2&F\\J&I&J\end{Bmatrix}\alpha^{(2)}_{nJ}. 
\end{align}
Here $I=9/2$ is the $^{115}\mathrm{In}$ nuclear spin. The quantity $\alpha^{(K)}_{nJ}$ is the reduced dynamical polarizability for state $\ket{nJ}$, where $n$ is the principal quantum number, and $K$ is the rank of the polarizability (i.e., $K=0$ for the scalar polarizability, $K=1$ for the vector, and $K=2$ for the tensor). The reduced polarizability is
\begin{widetext}
\begin{align}
    \alpha^{(K)}_{nJ} &= (-1)^{K+J+1}\sqrt{2K+1}\sum_{n'J'}(-1)^{J'}\begin{Bmatrix}
        1&K&1\\J&J'&J
    \end{Bmatrix}
    \abs{\bra{n'J'}\!\abs{\mathbf{d}}\!\ket{nJ}}^2\notag\\
    &\times\frac{1}{\hbar}\mathrm{Re}\,\qty(\frac{1}{\omega_{n'J'nJ}-\omega-i\gamma_{n'J'nJ}/2} + \frac{(-1)^K}{\omega_{n'J'nJ}+\omega+i\gamma_{n'J'nJ}/2}). \label{eq-pol-detailed}
\end{align}
\end{widetext}
In Eqn.~\ref{eq-pol-detailed}, $\hbar\omega_{n'J'nJ}$ is the energy difference between state $\ket{n'J'}$ and state $\ket{nJ}$, and $\gamma_{n'J'nJ}$ is the spontaneous decay rate from state $\ket{n'J'}$ to state $\ket{nJ}$. We obtain the reduced matrix element $\abs{\bra{n'J'}\!\abs{\mathbf{d}}\!\ket{nJ}}^2$ using
\begin{equation}
    \gamma_{n'J'nJ}=\frac{\omega_{n'J'nJ}^3}{3\pi\epsilon_0\hbar c^3}\frac{1}{2J'+1}\abs{\bra{n'J'}\!\abs{\mathbf{d}}\!\ket{nJ}}^2. 
\end{equation}
\begin{table*}
    \centering
    \caption{Data used to evaluate the polarizability. Numbers in brackets represent powers of $10$.}\label{tab:pol_raw_data}
    \begin{tabular}{*6c}\toprule
      \makecell{Lower\\state}  & \makecell{Upper\\state} & \makecell{Transition frequency\\($\mathrm{THz}$)} & \makecell{Frequency\\source} & \makecell{Decay rate\\($\mathrm{s}^{-1}$)} & \makecell{Decay rate\\source} \\ \midrule
$5^2P_{1/2}$ & $6^2S_{1/2}$ & $\phantom{0}730.9$ & \cite{nist-database} & $4.96[7]$ & \cite{nist-database} \\
$5^2P_{1/2}$ & $5^2D_{3/2}$ & $\phantom{0}986.4$ & \cite{nist-database} & $1.11[8]$ & \cite{nist-database} \\
$5^2P_{1/2}$ & $5^4P_{1/2}$ & $1048.9$ & \cite{nist-database} & $6.53[5]$ & \cite{nist-database} \\
$5^2P_{1/2}$ & $5^4P_{3/2}$ & $1080.2$ & \cite{nist-database} & $2.60[5]$ & \cite{nist-database} \\
$5^2P_{1/2}$ & $7^2S_{1/2}$ & $1088.6$ & \cite{nist-database} & $1.30[7]$ & \cite{nist-database} \\
$5^2P_{1/2}$ & $6^2D_{3/2}$ & $1171.0$ & \cite{nist-database} & $1.97[7]$ & \cite{nist-database} \\
$5^2P_{1/2}$ & $8^2S_{1/2}$ & $1218.6$ & \cite{nist-database} & $5.51[6]$ & \cite{nist-database} \\
$5^2P_{1/2}$ & $7^2D_{3/2}$ & $1254.6$ & \cite{nist-database} & $2.99[6]$ & \cite{nist-database} \\
$5^2P_{1/2}$ & $9^2S_{1/2}$ & $1281.1$ & \cite{nist-database} & $3.04[6]$ & \cite{nist-database} \\
$5^2P_{1/2}$ & $8^2D_{3/2}$ & $1299.6$ & \cite{nist-database} & $1.41[5]$ & \cite{nist-database} \\
$5^2P_{1/2}$ & $10^2S_{1/2}$ & $1315.9$ & \cite{nist-database} & $1.81[6]$ & \cite{nist-database} \\
$5^2P_{1/2}$ & $11^2S_{1/2}$ & $1337.4$ & \cite{nist-database} & $1.14[6]$ & \cite{nist-database} \\
$5^2P_{3/2}$ & $6^2S_{1/2}$ & $\phantom{0}664.5$ & \cite{nist-database} & $8.93[7]$ & \cite{nist-database} \\
$5^2P_{3/2}$ & $5^2D_{3/2}$ & $\phantom{0}920.0$ & \cite{nist-database} & $3.00[7]$ & \cite{nist-database} \\
$5^2P_{3/2}$ & $5^2D_{5/2}$ & $\phantom{0}920.7$ & \cite{nist-database} & $1.30[8]$ & \cite{nist-database} \\
$5^2P_{3/2}$ & $5^4P_{3/2}$ & $1013.8$ & \cite{nist-database} & $3.43[5]$ & \cite{nist-database} \\
$5^2P_{3/2}$ & $7^2S_{1/2}$ & $1022.3$ & \cite{nist-database} & $2.29[7]$ & \cite{nist-database} \\
$5^2P_{3/2}$ & $5^4P_{5/2}$ & $1056.8$ & \cite{nist-database} & $1.45[5]$ & \cite{nist-database} \\
$5^2P_{3/2}$ & $6^2D_{3/2}$ & $1104.6$ & \cite{nist-database} & $4.63[6]$ & \cite{nist-database} \\
$5^2P_{3/2}$ & $6^2D_{5/2}$ & $1106.1$ & \cite{nist-database} & $2.72[7]$ & \cite{nist-database} \\
$5^2P_{3/2}$ & $8^2S_{1/2}$ & $1152.3$ & \cite{nist-database} & $9.85[6]$ & \cite{nist-database} \\
$5^2P_{3/2}$ & $7^2D_{3/2}$ & $1204.1$ & \cite{nist-database} & $1.27[6]$ & \cite{nist-database} \\
$5^2P_{3/2}$ & $7^2D_{5/2}$ & $1189,0$ & \cite{nist-database} & $5.33[6]$ & \cite{nist-database} \\
$5^2P_{3/2}$ & $9^2S_{1/2}$ & $1214.7$ & \cite{nist-database} & $5.21[6]$ & \cite{nist-database} \\
$5^2P_{3/2}$ & $8^2D_{5/2}$ & $1233.8$ & \cite{nist-database} & $7.83[5]$ & \cite{nist-database} \\
$5^2P_{3/2}$ & $10^2S_{1/2}$ & $1249.6$ & \cite{nist-database} & $3.22[6]$ & \cite{nist-database} \\
$5^2P_{3/2}$ & $11^2S_{1/2}$ & $1271.0$ & \cite{nist-database} & $2.01[6]$ & \cite{nist-database} \\
$6^2S_{1/2}$ & $6^2P_{1/2}$ & $\phantom{0}223.2$ & \cite{nist-database} & $1.43[7]$ & \cite{Safronova2007} \\
$6^2S_{1/2}$ & $6^2P_{3/2}$ & $\phantom{0}232.1$ & \cite{nist-database} & $1.57[7]$ & \cite{Safronova2007} \\
$6^2S_{1/2}$ & $7^2P_{1/2}$ & $\phantom{0}434.3$ & \cite{nist-database} & $1.40[6]$ & \cite{Safronova2007} \\
$6^2S_{1/2}$ & $7^2P_{3/2}$ & $\phantom{0}437.7$ & \cite{nist-database} & $1.96[6]$ & \cite{Safronova2007} \\
$6^2S_{1/2}$ & $8^2P_{1/2}$ & $\phantom{0}523.3$ & \cite{nist-database} & $4.07[5]$ & \cite{Safronova2007} \\
$6^2S_{1/2}$ & $8^2P_{3/2}$ & $\phantom{0}524.9$ & \cite{nist-database} & $6.40[5]$ & \cite{Safronova2007} \\
$6^2P_{3/2}$ & $5^2D_{5/2}$ & $\phantom{0}\phantom{0}24.0$ & \cite{nist-database} & $1.03[5]$ & \cite{Safronova2007} \\
$5^2D_{5/2}$ & $7^2P_{3/2}$ & $\phantom{0}181.6$ & \cite{nist-database} & $5.67[5]$ & \cite{Safronova2007} \\
$5^2D_{5/2}$ & $4^2F_{5/2}$ & $\phantom{0}203.6$ & \cite{nist-database} & $9.46[5]$ & \cite{Safronova2007} \\
$5^2D_{5/2}$ & $4^2F_{7/2}$ & $\phantom{0}203.6$ & \cite{nist-database} & $1.42[7]$ & \cite{Safronova2007} \\
$5^2D_{5/2}$ & $8^2P_{3/2}$ & $\phantom{0}268.8$ & \cite{nist-database} & $2.29[5]$ & \cite{Safronova2007} \\
$5^2D_{5/2}$ & $5^2F_{5/2}$ & $\phantom{0}280.0$ & \cite{nist-database} & $3.93[5]$ & \cite{Safronova2007} \\
$5^2D_{5/2}$ & $5^2F_{7/2}$ & $\phantom{0}280.0$ & \cite{nist-database} & $5.90[6]$ & \cite{Safronova2007} \\
 \bottomrule
    \end{tabular}
    \label{tab:transition_data}
\end{table*}

The atomic transition data used in the polarizability calculation is shown in Table \ref{tab:transition_data}. All data for $\omega_{n'J'nJ}$ are experimental values taken from the NIST Spectral Database. We use the experimental transition rates that are available on the NIST Spectral Database, and when experimental values are not available, we use theoretical ones from Ref~\cite{Safronova2007}. We also use a theoretical value of the indium core polarizability~\cite{Safronova2013}. 

With these values and the above-mentioned equations, we calculate the polarizabilities of the ground state $\ket{5P_{1/2}, F=4}$, the metastable lower cooling state $\ket{5P_{3/2}, F=6}$, the upper repumper state $\ket{6S_{1/2}, F = 5}$, and the upper cooling state $\ket{5D_{5/2}, F = 7}$, as shown in Fig. \ref*{fig-ol} of the main text. 

\subsection{Loading into an optical lattice}

\begin{figure}
    \includegraphics[width=.5\linewidth]{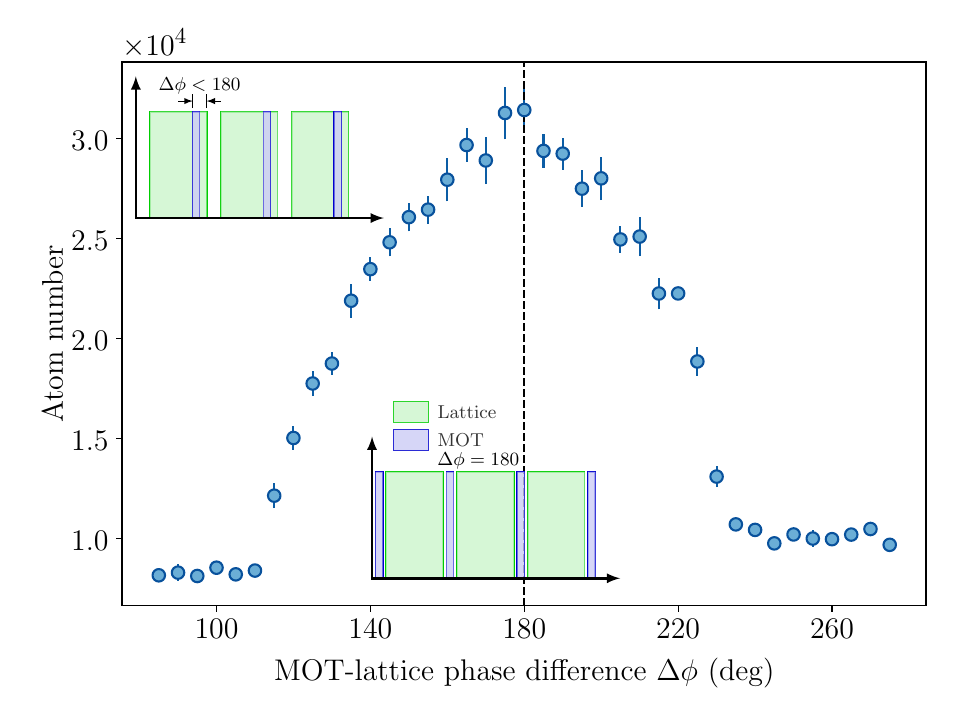}
    \caption{Loading efficiency as a function of the phase difference between cooling and trapping pulses. Optimal loading efficiency occurs at $\Delta\phi=180^\circ$.}\label{fig-ol-phase}
\end{figure}

In addition to the intensity modulation data of Fig.~\ref*{fig-ol} in the main text, we also studied the effect of the MOT-lattice pulse phase difference on the trapping efficiency (Fig.~\ref{fig-ol-phase}). In other experiments using this modulation technique, the optimum phase difference was observed to be $180^{\circ}$~\cite{Aliyu2021}; however, one attempt significantly deviated from $180^{\circ}$~\cite{Hutzler2017}. In our system, we observe that $180^{\circ}$ is optimal. 

We estimate the phase space density after lattice loading to be $4\times 10^{-2}$. This occurs at a density of $n = 2.4\times10^{15}\,\mathrm{cm}^{-3}$ and a trap beam waist of $w_0=33\,\mathrm{\mu m}$. The axial trap frequency for atoms in $\ket{5P_{1/2}, F=6, m_F=0}$ is $\omega_a=2\pi\times 247\,\mathrm{kHz}$ at the beam focus. This value is accurate up to 2 mm away from the center, making it valid for the entire trapped gas.

\subsection{Optical lattice density data}

\begin{figure}[h!]
\parbox{.495\textwidth}{
\includegraphics[width=.475\textwidth]{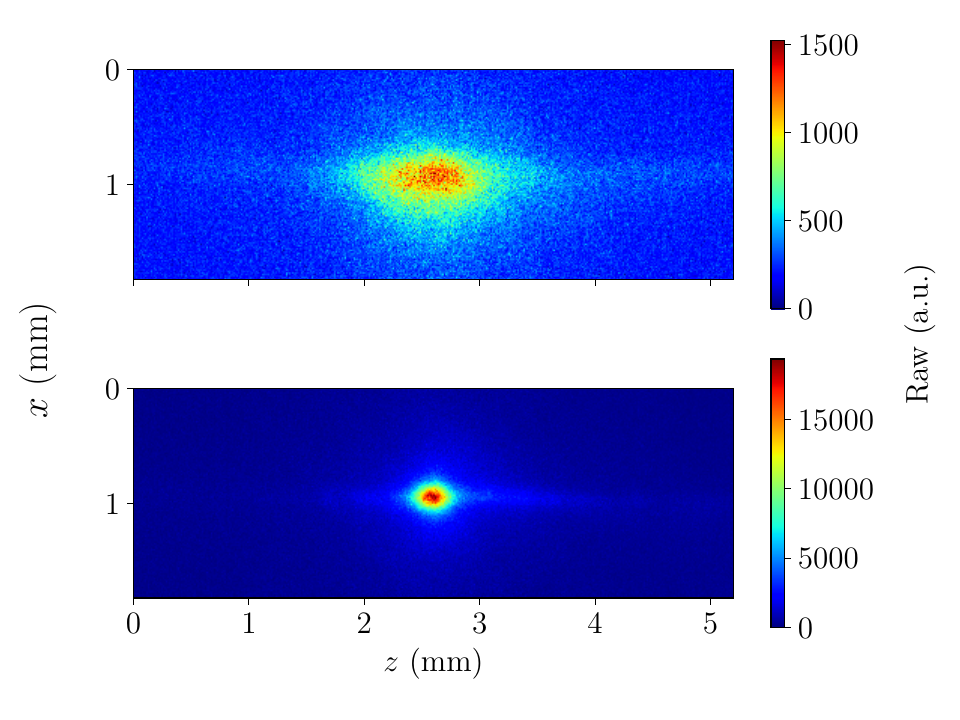}
\subcaption{ Fluorescence images after 2.5 ms free expansion.
TOP: Atoms released from an optical dipole trap (ODT). The ODT is achieved by blocking the retroreflecting lattice beam. BOTTOM: Atoms released from an optical lattice, which forms when the retroreflecting beam is no longer blocked.}\label{fig-lattice-fluo-a}
}\hspace{2pt}
\parbox{.495\textwidth}{
\includegraphics[width=.475\textwidth]{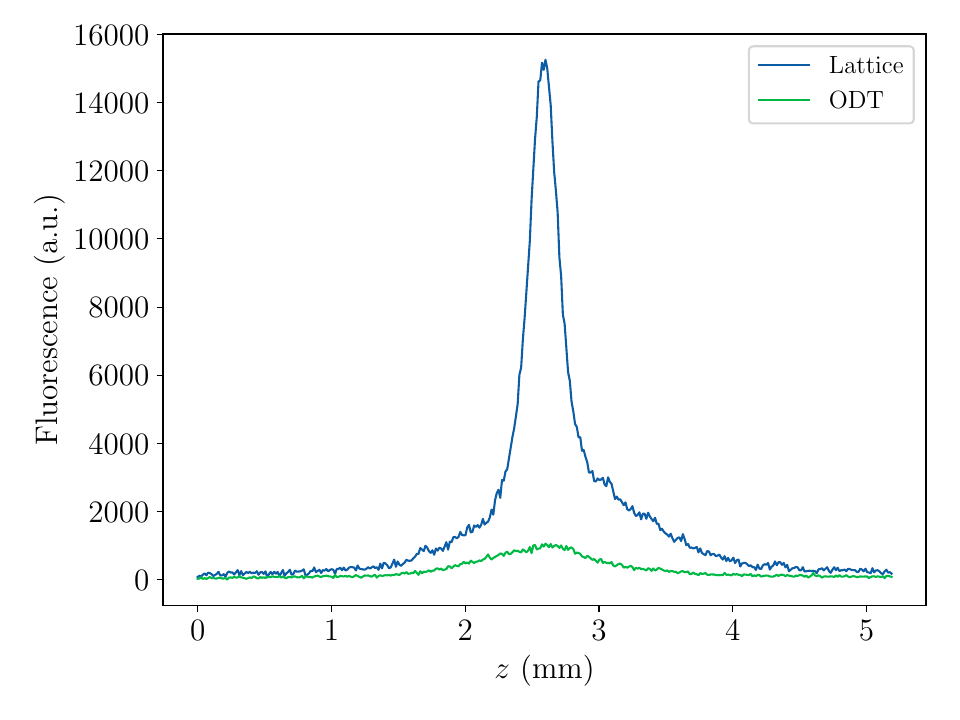}
\subcaption{Line density profiles obtained by summing the fluorescence images in (\subref{fig-lattice-fluo-a}) over the transverse ($x$) axis.
The strong central peak occurs due to the axial confinement from the lattice.
}
}
\caption{Comparison of atomic density distributions after free expansion from an optical dipole trap and an optical lattice.
}\label{fig-fitted-lattice-density}
\end{figure}

Our atom number data comes from atomic density measurements (Figure \ref{fig-lattice-fluo-a}). To better understand the density data, we developed a numerical model to simulate the density. This model treats the electric field of the lattice as
\begin{equation}
\mathbf{E}(x,y,z,t) = \hat{x} e^{-i \omega t} \left[
E_1 \frac{w_{0,1}}{w_1(z)} e^{- \frac{x^2 + y^2}{w^2_1(z)}} e^{i k z} +
E_2 \frac{w_{0,2}}{w_2(z)} e^{- \frac{x^2 + y^2}{w^2_2(z)}} e^{- i k z}
\right],
\end{equation}
where $k=2\pi/\lambda$ is the wavevector, $E_{i} \propto \sqrt{P_i}$ ($i=1,2$), $w_{0,i}$ is the beam waist at the focus, $w_i(z) = w_{0,i} \sqrt{1 + (z/z_{R,i})^2}$ is the beam waist at axial position $z$, and $z_{R,i} = \pi w_i^2 / \lambda$ the Rayleigh range. Here the $E_1$ term is the incident beam and the $E_2$ term is the retroreflected beam. We neglect terms due to wavefront curvature and Gouy phase. This models a trap field polarized in the $\hat{x}$ direction that can have different values for the incident and reflected powers and focal waists. 

The trap potential is proportional to $\left| \mathbf{E}(x,y,z,t) \right|^2$. The simulation treats the lattice as being loaded from a MOT. MOT loading is modeled by drawing a random velocity for each of the 3 spatial dimensions from a Maxwell-Boltzmann distribution, with a temperature equal to our measured MOT temperature. The position is also drawn from a Gaussian function with dimensions equal to our measured MOT size. At the resulting position values, the potential energy from the lattice is calculated. If the kinetic plus potential energies are greater than the trap depth, the atom is treated as unsuccessfully trapped and removed from the simulation. 

\begin{figure}
    \includegraphics[width=.5\textwidth]{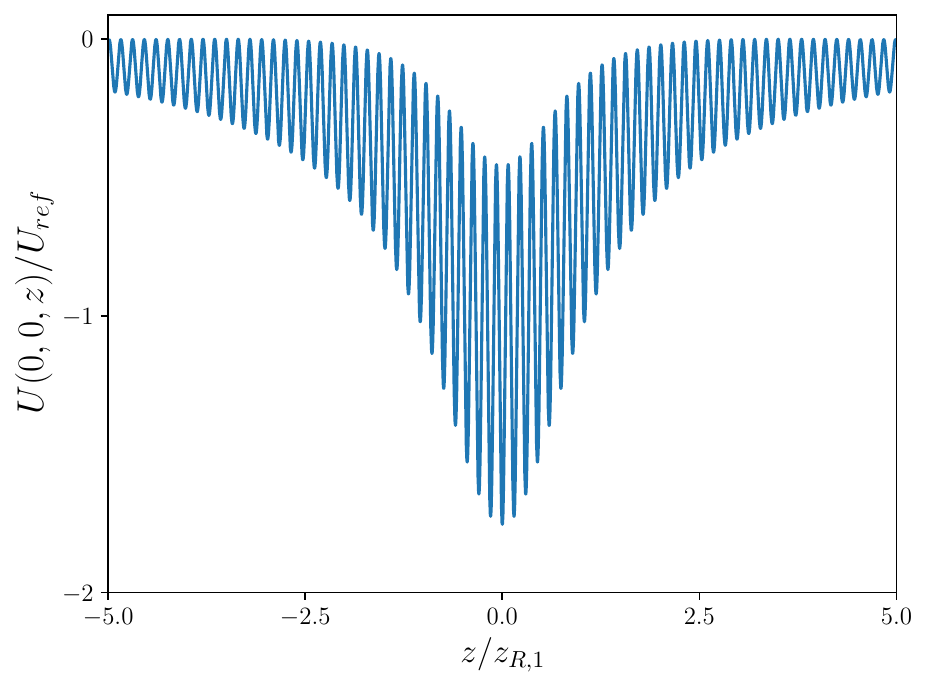}
    \caption{
    Lattice potential for the measured value of $|E_2/E_1| = \sqrt{0.556}$ and the simulation-determined value $w_{0,2}/w_{0,1} = 2.3$. For the purpose of visualizing the lattice sites, the laser frequency in the plot is 30 times smaller than the experimental value. The potential is proportional to the squared magnitude of the electric field, $|\vec{E}|^2$, and the reference potential $U_{ref}$ is the peak potential depth of the incident beam (retroreflected beam is blocked). Near zero potential depth, imperfect interference contrast creates a broad background potential, which is an extended region of shallow axial confinement at energies between the lattice sites and 0. These amplitude and waist ratios modify the depth and curvature of the confining envelope.}\label{fig-theory-lat-potential}
\end{figure}

Modeling the lattice as an incident laser with an imperfectly matched retroreflected beam results in a potential that has lattice sites at the bottom of a broader background potential (Fig.~\ref{fig-theory-lat-potential}). Atoms that spawn with total energies below the depth of the nearest lattice site are treated as frozen in that site. Freezing them in accounts for quantized motion. Atoms spawning with energy above this depth can freely explore the background potential.

Atoms confined to lattice sites are assumed to remain stationary, but atoms in the background potential must be simulated dynamically due to their mobility. To model their phase space evolution, we use the Velocity-Verlet method, which integrates Newton’s equations of motion. At each time step, the local acceleration (determined by the gradient of the potential) is treated as approximately constant, allowing for position and velocity updates via standard kinematic relations~\cite{Verlet1967}.

\begin{figure}
    \includegraphics[width=.5\textwidth]{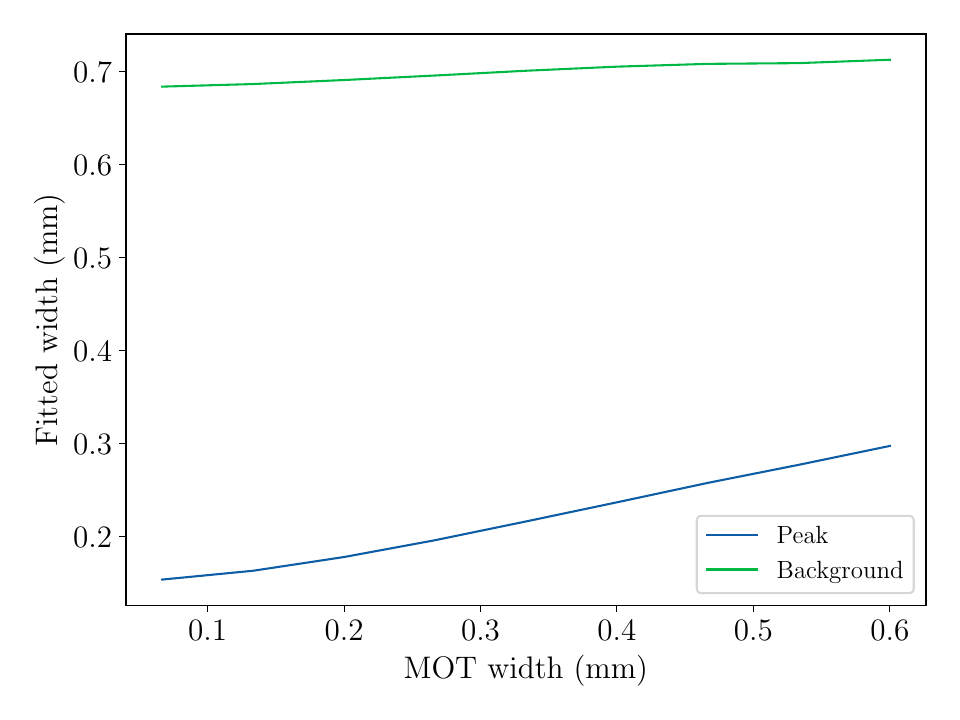}
\caption{Dependence of fitted widths on initial MOT size. The width of the background component grows slowly, while the central peak widens significantly as the initial MOT size increases.}\label{fig-width-dependency}
\end{figure}

The result is a simulated density with a sharp central peak overlaid on a broader background feature (Figure \ref{fig-theory-lat-potential}). The central peak corresponds to the density of particles in the lattice sites, whereas the broad feature comes from the hotter atoms in the background potential. We find that the central peak is sensitive to the width of the MOT, whereas the background feature is not (Fig. \ref{fig-width-dependency}). This is a result of the lattice's stronger axial confinement, which is a qualitative difference between a lattice and an optical dipole trap (ODT). Meanwhile atoms in the background potential relax along the axial direction from their MOT distribution into a larger one determined by the ODT potential shape.

Power in the retroreflected beam is lost due to absorption in the viewports. We measure a power loss coefficient of 74.5\% for a single viewport, which is consistent with the known transmission dip at 1064 nm in UV fused silica (the material our viewports are made of). Using this, we calculate a power in the incident lattice beam of $P_1 = 33 \ \mathrm{W}$ at the atoms. Due to the retroreflected beam double passing one of these viewports, the reflected beam power is reduced by 55.6\% when it reaches the atoms.

\begin{figure}
\includegraphics[width=.5\textwidth]{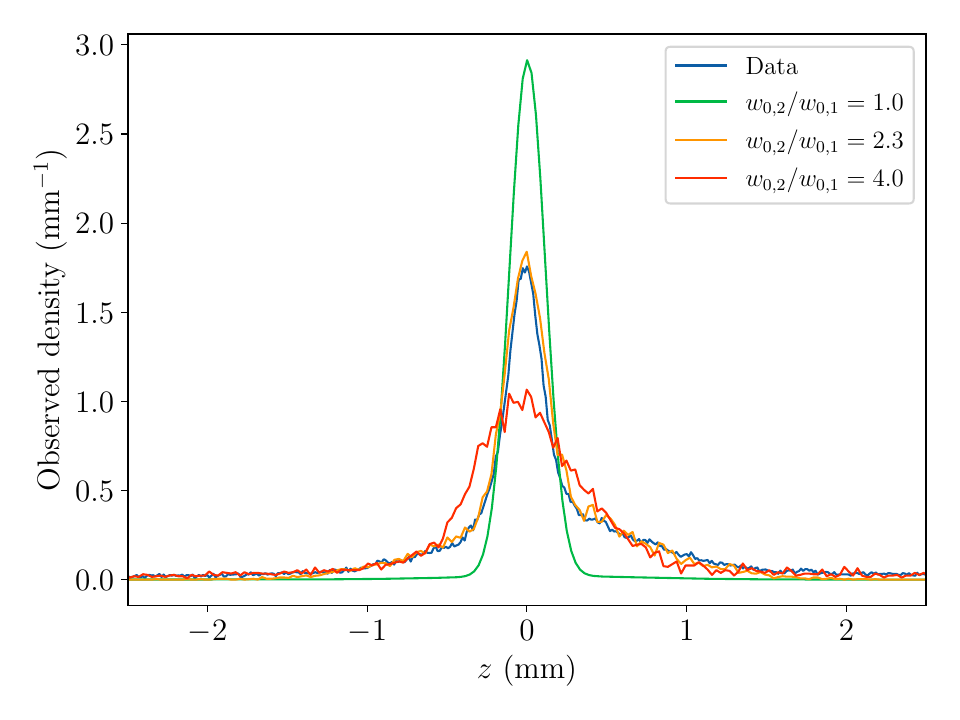}
\caption{Simulated axial density distributions for different retroreflected beam waist ratios $w_{0,2}/w_{0,1}$, compared with experimental data. The simulation, performed with the velocity-Verlet method under the same trapping conditions, reproduces both the sharp central peak and the broad background wings. Broadening of density profile due to the 2.5 ms free expansion time used in the experiment is included in the simulated curves. Best agreement with the data is obtained at about $w_{0,2}/w_{0,1}=2.3$.}\label{fig-density-theory}
\end{figure}

We also consider the case of a reflected beam with a waist that is imperfectly matched to the incident beam. We measure the waist of the incident beam to be $w_{0,1} = 33 \ \mathrm{\mu m}$. By varying the retroreflected beam waist $w_{0,2}$, we find that the simulation agrees well with the data when $w_{0,2}/w_{0,1} = 2.3$ (Fig. \ref{fig-density-theory}). Although this power attenuation and waist ratio are needed for agreement with the atoms in the background potential, these have little effect on the width of the sharp peak. 

We note that the 2.5 ms free expansion time that occurs before the density is imaged is built into the model. With this time included, we find that the width of the sharp peak in the simulation agrees with our data when we use an accurate model of the MOT density. 

\subsection{Measurement and fitting of the decay models}

For our measurements of in-trap decay of both $5P_{1/2}$ and $5P_{3/2}$, the procedure for initializing the state in the optical lattice is as follows. After intensity-modulated loading of the lattice, we switch on the necessary repumpers to pump atoms into either the $\ket{5P_{1/2},F = 4}$ or $\ket{5P_{3/2},F = 6}$ states. During this pumping sequence, the lattice is shut off to avoid AC Stark shifts on the 410 and 451 nm transitions. The lattice only needs to be shut off for $70\ \unit{\mu s}$ to complete the pumping phase. After this, atoms are held in the lattice for a variable length of time, pumped into $\ket{5P_{3/2},F = 6}$ (for the case of the $\ket{5P_{1/2},F = 4}$ data), and then imaged using fluorescence from the cooling transition. 

For pure one-body decay, the decay rate is
\begin{equation}
    \dv{N}{t}=-\frac{N}{\tau},  
\end{equation}
where $\tau$ is the one-body decay time. This results in the familiar solution
\begin{equation}
    N(t) = N(0) e^{-t/\tau}.
\end{equation}
If both one- and two-body decay are present, particle loss is governed by the differential equation
\begin{equation}
    \dv{N}{t} = - \frac{N}{\tau} - \Gamma_2 N^2.
    \label{eq: differential equation}
\end{equation}
Here $\Gamma_2 = \beta/V$ is the two-body decay rate, $\beta$ is the two-body loss coefficient, and $V$ is the trap volume. Also $\sigma_x$, $\sigma_y$, and $\sigma_z$ are the RMS radii of the trapped cloud. The solution to this differential equation is
\begin{equation}
    N(t)=\frac{N_0e^{-t/\tau}}{1+N_0\Gamma_2\tau(1-e^{-t/\tau})}. \label{eq-ol-two-body-decay}
\end{equation}

We fit the two-body solution to decay data from the $\ket{5P_{3/2},F=6}$ state and find that
\begin{align}
    \tau &= 3.0(1)\,\mathrm{s} \\
    \Gamma_2 &= 8.4(3)\times10^{-5}\,\mathrm{s}^{-1}.
\end{align}
For decay from the ground state, we perform a statistical $F$-test to determine whether the two-body model is necessary or whether we can use the one-body model only. For 21 observations, we calculate a reduced $\chi^2$ of 3.4 for the one-body model and 2.0 for the two-body case. This results in an $F$ score of 12.5, corresponding to a $p$ value of $2.2\times10^{-3}$. We therefore conclude that two-body decay cannot be ignored in the ground state. The two-body fit of the ground state data yields
\begin{align}
    \tau &= 2.7(2)\,\mathrm{s} \\
    \Gamma_2 &= 1.4(3)\times10^{-5}\,\mathrm{s}^{-1}.
\end{align}
Note here that the one-body loss time is statistically consistent across both measurements; however, the two-body loss rate is 6 times lower in the ground state.

\bibliography{supplemental.bib}